\documentclass[12pt,letterpaper]{article}
\usepackage[a4paper, total={7in, 10in}]{geometry}
\renewcommand{\familydefault}{\sfdefault}
\usepackage{graphicx}
\usepackage{helvet}
\usepackage{authblk}
\usepackage{hyperref}
\usepackage{amsmath} 
\usepackage{color,soul}
\soulregister\cite7
\soulregister\ref7
\soulregister\cref7
\soulregister\pageref7
\soulregister\url7
\soulregister\item7
\usepackage{amssymb} 
\usepackage[super,comma,sort&compress]  
   {natbib}
\usepackage[right]{lineno} \linenumbers

\usepackage{float}
\usepackage{xcolor}
\usepackage{epstopdf}

\usepackage{xr}
\usepackage{zref-xr}
            \makeatletter
            \newcommand*{\addFileDependency}[1]{
              \typeout{(#1)}
              \@addtofilelist{#1}
              \IfFileExists{#1}{}{\typeout{No file #1.}}
            }
            \makeatother

\graphicspath{
    {figures}
}
\usepackage{cleveref}
\newcounter{vid}
\crefname{vid}{S}{S}
\crefformat{vid}{S~#2#1#3}

\AtBeginDocument{%
  \crefformat{vid}{#2S#1#3}%
}
\usepackage{ragged2e}

\makeatletter
\renewcommand{\maketitle}{\bgroup\setlength{\parindent}{0pt}
\begin{flushleft}
  \textbf{\@title}
  
  \@author
\end{flushleft}\egroup}
\makeatother

\title{Collective action and entanglement of magnetically active liquid crystal elastomer ribbons}
\date{}

\author[1,2,5]{Asaf Dana}
\author[1]{Christian Benson}
\author[1]{Manivannan Sivaperuman Kalairaj}
\author[1]{Kayla Hellikson}
\author[1]{Sasha M. George}
\author[1]{David C. Chimene}
\author[1]{Jared A. Gibson}
\author[1]{Seelay Tasmim}
\author[3]{Phillip A. Kohl}
\author[3]{Youli Li}
\author[2]{Mustafa K. Abdelrahman}
\author[4]{Vishal P. Patil}
\author[1,2,5,6,*]{Taylor H. Ware}

\affil[1]{Department of Biomedical Engineering, Texas A\&M University, College Station, TX 77843, USA}
\affil[2]{Department of Materials Science and Engineering, Texas A\&M University, College Station, TX 77843, USA}
\affil[3]{Materials Research Laboratory, University of California - Santa Barbara, Santa Barbara, CA 93106, USA}
\affil[4]{Department of Mathematics, University of California, San Diego, La Jolla, CA 92093, USA}
\affil[5]{Senior author}
\affil[6]{Lead contact}

\affil[*]{Correspondence: taylor.ware@tamu.edu}

\begin{document}

\maketitle

\section*{SUMMARY}

Interactions between active individuals in animal collectives lead to emergent responses that remain elusive in synthetic soft matter.
Here, shape-morphing polymers serve to create bio-inspired transient solids that self-assemble with controlled mechanical properties and disassemble on demand. 
Dilute-suspensions of magnetic, heat-responsive liquid crystal elastomer ribbons mechanically interlock, inducing reversible aggregation. 
A mathematical model is developed that sheds light on the role of topological mechanisms in aggregation.
Aggregation was favored for ribbons with moderate curvature as compared to flat ribbons or higher curvature ribbons.
The ribbon suspensions reversibly transition between fluid- and solid-like states, exhibiting up to 6 orders-of-magnitude increase in the storage moduli of the entangled aggregates compared with the liquid dispersions. 
Controlled dissociation is induced by imparting kinetic energy to the individual ribbons at high magnetic field rotation speeds ($>3.33$ Hz). 
Imparting dynamic collective behaviors into synthetic systems may enable potential applications from bio-inspired soft robotics to injectable biomaterials.

\section*{KEYWORDS}


liquid crystal elastomers, active matter, collective action, entangled matter, actuators, stimuli-responsive polymers, soft robotics

\section*{INTRODUCTION}

\indent

Materials capable of self-assembly and disassembly in response to external cues hold promise to enable new functionality across engineering disciplines \cite{Fan2020,Yang2022,Diller2013}.
Such autonomous behaviors may offer novelty in fabrication and processing practices of bulk functional material, as well as in soft robotics, in which a swarming behavior of individual units can be manipulated to result in a stable solid structure \cite{Diller2013,Saintyves2024}. 
If such structures do not posses long range order (amorphous), they may further hold the benefit of requiring low levels of control over the assembly process to produce a multitude of forms with consistent properties \cite{Savoie2023,Abdelrahman2024}.

One interesting approach towards building materials that self-assemble and disassemble is to rely on interactions between units that are able to move independently.
The general term used for such systems is active matter.
Such systems are composed of individual active units (across a variety of scales) that are far from equilibrium, which consume energy to generate motion or mechanical work \cite{Ramaswamy2010,Vernerey2019}.
In many cases, the propelling  of individual elements drives the self-assembly of the units into a structure or aggregate leading to collective behavior and emergent properties.
If the cohesive forces between the aggregate elements are a result of the continuous supply of energy to the elements, the structure will break down upon consuming the source or the cutoff of the energy supply.
In some cases, collisions may result in jamming or entanglement which can lead to metastable assembly due to the frustrated interaction between the units \cite{Behringer2018,Vernerey2019}.

Some interesting sources of inspiration to cohesive interactions between active individuals that lead to collective mechanical behaviors can be found in the animal kingdom \cite{Vernerey2019}.
Notable examples are fire ants (\textit{Solenopsis invicta}) \cite{Wagner2021}, honeybees (\textit{Apis mellifera}) \cite{Grozinger2014}, plant–animal worms (\textit{Symsagittifera roscoffensis}) \cite{Worley2019}, and California blackworms (\textit{Lumbriculus variegatus}) \cite{Patil2023}.
These species mechanically interlock (in response to environmental conditions) into cohesive masses that manifest various collective behaviors such as buoyant raft formation \cite{Mlot2011}, colony unification \cite{Anderson2002} and thermal regulation \cite{Tennenbaum2016}. 
By nature, these masses are porous structures with viscoelastic mechanical behaviors \cite{Ozkan2021} that exhibit adaptive self-assembly and reorganization.
These properties stem from each member’s ability to move within the confines of the structure, or in-and-out by physically detaching and reattaching to its neighbours \cite{Anderson2002,Foster2014,Tennenbaum2017,Deblais2020}.
Moreover, their ability to reversibly transition between fluid- and solid-like states allows them to be uniquely adaptable and, as a mechanical system, to operate with a large number of degrees of freedom \cite{Aguilar2016review,Wagner2022}.
In the context of a  robotic system or a synthetic functional device made with a clear engineering application in mind, such capabilities have only been partially leveraged.

To-date, several synthetic realizations of interlocked aggregates have been realized.
For example, a variety of granular and swarming robotic systems that self-assemble into two dimensional aggregates using loose magnetic coupling have been proposed.
These systems were able to demonstrate transitions between collective states with solid- and liquid-like properties \cite{Saintyves2024} and exhibited emergent collective properties such as locomotion and transport \cite{Li2021,Li2019}. 
However, this approach uses bulky robotic components, each requiring relatively complex integrated electronic circuits and controls.
Other approaches employ the unique attributes of responsive polymeric materials.
The use of polymeric materials can greatly enhance the simplicity of the responsive unit.
 {Wie et al., \cite{Won2022,Yang2024} demonstrated collective action of magnetic assemblies of anisotropic composite polymer robots.
They have shown that the exertion of kinetic energy through controlled rotation of magnetic fields induces swarming behaviors that can operate in both dry and wet environments.
These swarms further demonstrated versatile task execution such as controlled assembly, disassembly, swimming, stirring and object manipulation. }
Abdelrahman et al. \cite{Abdelrahman2024} demonstrated the creation of transient viscoelastic aggregates from a range of functional polymer ribbons.
Such materials included liquid crystal elastomers (LCE) which are loosely crosslinked responsive polymer networks that reversibly change their shape in response to external stimuli such as heat or light.
However, the aggregation required a temperature transient and large concentrations of overlapping entangling units for aggregate formation. 
 {A combination  of the latter concepts utilizing functional polymers can offer a versatile platform for the creation of multifunctional programmable materials that exhibit a wide array of emergent collective behaviors.}

In this work, we describe an approach to create transient, responsive solids from dilute suspension of mobile shape-changing polymeric ribbons.
Specifically, we utilize the large-scale motion (i.e., much greater than a single body length) exhibited by magnetic and shape changing LCE ribbons in response to a rotating uniform magnetic field.
When heated, ribbons change shape with prescribed amounts of bending and twisting, Figure \ref{fig:boat1}a.
Then, the rotation of the magnetic field induces aggregation through the entanglement of individual ribbons together into clusters, e.g., Figure \ref{fig:boat1}b and Supplementary Video \cref{vid:S1}.
We describe the thermal and magnetic response of individual ribbons, and track the aggregation kinetics of ribbon dispersions.
Following the rotation of the magnetic field at elevated temperatures (65 $^\circ$C, Figure \ref{fig:boat1}c), the aggregates  have a greater storage modulus than loss modulus up to shear strains of approximately 3\%, constituting a solid-like behaviour, Figure \ref{fig:boat1}d.
The viscoelastic mechanical and rheological properties of the formed aggregates are found to vary with the single unit shape.
Finally, we examine the critical conditions required for controlled aggregate disassembly.

\section*{RESULTS}

\subsection*{Thermal and Magnetic response\label{sec_thermal_response}}

LCE ribbons that respond to heat by changing shape and respond by moving in an external magnetic field were fabricated. 
The response to the magnetic field is conferred via a composite LCE domain at the edge of the ribbon filled with magnetic particles, Figure \ref{fig:boat1}a and Supplementary Figure S1. 
The remainder of the LCE ribbon is shape responsive. 
The nematic director rotates by 90 $^\circ$ through the film thickness (twisted nematic, Figure \ref{fig:boat2}a), and heating causes the ribbon to morph from flat to 3D shapes with controlled degrees of bend and twist \cite{Sawa2013,De2014}.
Grazing incidence WAXS measurement on monodomain nematic film samples (uniform through the thickness) validated a high degree of molecular alignment along the prescribed orientation (Figure S1).
The thermally induced transition from the nematic phase towards the isotropic phase locally induces a contraction along the nematic director that is accompanied by an expansion in the perpendicular direction.  
The amount of contraction is prescribed by the change in temperature. 
 {The most shape change occurs when heating through the nematic to isotropic transition temperature, noted as $T_{NI}$. In these materials some order is always retained on heating as such this transition is sometimes called the nematic-paranematic transition\cite{Verwey1997}.}
Twisted nematic ribbons with the nematic director at each face offset by an angle $\theta$ to the principal axes of the ribbon, Figure \ref{fig:boat2}b, exhibit controlled bending and twisting \cite{Sawa2013,De2014}.
The evolution of curvature with temperature for ribbons cut at different offset angles is shown in Figure \ref{fig:boat2}d, and the amount of twist (around the long axis of the ribbon) per unit length at different temperature is shown in Figure \ref{fig:boat2}e. 
Ribbons that were not cut at an offset angle, $\theta = 0^\circ$, only bend, show an approximately linear rate of increase in curvature of approximately 0.014 mm$^{-1}$/$^\circ$C. As expected, increasing the offset angle $\theta$ shows a decrease in curvature and an increase in the amount of observed twist.
Mild offset angles of $\theta = 10^\circ$ showed similar curvatures to  $\theta = 0^\circ$ with an increased amount of twist. 
The twisting of the ribbon gives rise to a helical shape, Figure \ref{fig:boat2}b, with the pitch proportional to the amount of twist.
Specifically, for a 16 mm ribbon at 65 $^\circ$C, there is a quarter twist along the long axis of the ribbon.  
 {Close-up images of actuated ribbons at different temperatures and offset angles are presented in Supplementary Figure S2.}
As temperature increases towards $T_{NI}$, the curvature of the ribbon increases, Figure \ref{fig:boat1}a.
Thus, the coil becomes tighter with decreases in pitch, length, and diameter.

Once the ribbon has acquired its actuated shape under the prescribed temperature, rotation of an external magnetic field is applied to induce the motion of the ribbon.
 {The motion of the ribbon is a result of multiple factors including the magnitude of the magnetization in the ribbon head ($14.9\pm0.21$ emu/g, Supplementary Figure S1), the external magnetic field ($26$ mT, Supplementary Figure S3), the geometry of the ribbon, and the viscosity of the media.
The ratio of inertial force to viscous force acting on the rotating swimming ribbon in a viscous fluid can be estimated by the rotational form of the Reynolds number \cite{Han2021low}, $Re_\omega = \rho\omega r^2/\eta$.
Substitution of the parameters of the actuated ribbon at 65 $^\circ$C and the viscous fluid yields an approximate value of  $Re_\omega\approx 1.5$, indicating a close balance between the forces and a laminar flow regime.
Additional heating of the system will result in an increase in curvature (squared decrease in radius) and a proportional reduction in viscosity, indicating that this provides an upper bound for the relative magnitude of inertial forces. }
 {To evaluate the effect of rotation on the ribbon shape, we measured the change in curvature of individual ribbons in the presence of the rotating magnetic field.
Ribbons exhibited an average   $8.1\pm1.7$ \% reduction in curvature during  rotation at 1.2 Hz compared to the rest state ($n=3$).}

Magnetic actuation of different ribbon shapes, caused by heating to different temperatures, induces trajectories that vary in space and time (Supplementary Figure S4 and Supplementary Video \cref{vid:S2}).
To obtain a measure of the response of ribbons to an applied rotating field, the position of the magnetic domain was tracked in time in the $x-y$ plane (parallel to the fluid surface).
The collected trajectories were filtered to a small range around the magnetic field rotation frequency, i.e., approximately 1.2 Hz (72 RPM).

The measured actuation paths of the ribbons consist of repeatable patterns that indicate three distinct motion regimes.
The first pattern is characterized by many acute-angled features, 40 $^\circ$C in Figures \ref{fig:boat2}f-h. 
In this regime the ribbon frequently changes its direction of motion.
The magnetic domain proceeds in a side-to-side flapping-like motion (e.g., Supplementary Video \cref{vid:S2}).  
This pattern of motion typically occurs when the ribbon is flat, and the magnetic moment is unable to make the ribbon bend or rotate about its axis before the external field completes a full rotation.
The second motion regime is more circular, 65 $^\circ$C-120 $^\circ$C in Figures \ref{fig:boat2}g-i, indicating that on the time scale related to the magnetic field rotation, the ribbon is able to follow the field without changing directions (e.g., Supplementary Video \cref{vid:S2}).
The third motion regime contains more relaxed angular features oriented in different directions 65 $^\circ$C-120 $^\circ$C in Figure \ref{fig:boat2}f.
This pattern indicates a transition regime in which the ribbon is able to partially follow the field rotation but still displays a sizeable, albeit smaller, amount of direction changes per rotation cycle.

We hypothesize that the conditions required for successful entanglement in our system are twofold.
First, the ribbon should be able to successfully follow the field to maintain synchronicity with its cyclic trajectories.
Such trajectories are typically represented by the more circular patterns, 65$^\circ$C-120 $^\circ$C in Figures \ref{fig:boat2}g-i.
Second, the ribbon should have a moderate amount of curvature that will allow ribbons to physically interlock with each other but not restrict interaction.

\subsection*{Assembly and aggregation kinetics\label{sec_kinetics}}

Magnetic actuation of many curved ribbons in a fluid induces isothermal aggregate assembly from dispersed states. 
 {In each of the following experiments, forty ribbons were initially placed in a dish filled with viscous silicone oil, corresponding to a LCE packing density of 0.002 mg/mm$^3$ (or a volume fraction of  0.0011 mm$^3$/mm$^3$). }
The ribbons were placed in the oil such that there was no overlap between them ($t=0$, Figure \ref{fig:boat3}a).
 {The oil submersion allows for maintaining a constant temperature throughout the system, but it may also act to reduce adhesion between the ribbons, and supports motion in 3D by reducing effects of gravity. }
The motion induced by the magnetic field at low rotation speeds (1.2 Hz or 72 RPM) facilitated entanglement of pairs of individual ribbons. Pairs subsequently grew into clusters through the annexation of additional ribbons or ribbon clusters, Supplementary Video \cref{vid:S1}.
 {The side view in Supplementary Video \cref{vid:S1} shows that the process occurs on the bottom of the dish in a fully submersed state, such that the surface tension of the oil does not affect the process.}

 {Entanglement results from curved ribbons threading into each other.
The rotation of the ribbons in the dish is accompanied by collisions of rotating individual ribbons or ribbon clusters.
Such collisions result in topological interaction in which the active magnetic domain of one ribbon threads into the curved body of another and they begin to rotate together until interaction repeats with an additional ribbon and the cluster grows, Supplementary Video \cref{vid:S3} and Supplementary Video \cref{vid:S4}.
Although interaction of the magnetic domains can occur in specific orientations, the weak forces between the magnetic domains (Supplementary Figure S1) are not sufficiently strong to hold the structure together or withstand the forced motions induced by the rotating field in the absence of ribbon curvature.
This is further true for the cohesive interaction between the non-magnetic ribbon bodies, supporting the assumption that capillarity (e.g., Ref \cite{Hu2020}) does not play a dominant role in interaction.
To illustrate this, the entangled structure can be cooled so that the ribbons return to a flat shape, and the magnetic field rotation is subsequently turned on again. 
Despite the close proximity of the magnetic domains and the ribbon bodies to each other, cooling and restarting the magnetic field  results in a redispersion of the ribbons along the dish propelled by the previously described flapping motion (Supplementary Video \cref{vid:S3} and Supplementary Video \cref{vid:S4}).}
{These findings lead us to hypothesize that the main contribution for cohesive interaction stems from static friction between ribbon surfaces.
Additionally, we observe that ribbon clusters tend to rotate as a somewhat rigid body with minimal relative motion of ribbons within the cluster.}

Aggregation varies based on unit shape which is prescribed by the temperature. 
Time lapse images of dispersions of ribbons cut at ($\theta=10^\circ$)  at different temperatures show different aggregation behavior, Figure \ref{fig:boat3}a. 
At all temperatures that induce shape change, aggregation is observed. 
At a common actuation temperature of 65$^\circ$C, ribbons that bend and/or twist aggregate while those that remain flat (polydomain) do not.
The time evolution of the \# of clusters for offset angle $\theta=10^\circ$ at different temperatures ($T=[65,90,120]^\circ$C) is presented in Figure \ref{fig:boat3}b.
At 40 $^\circ$C (red line), the ribbons remained flat and exhibited discontinuous flapping motion, similar to the motion pattern exhibited by single ribbons in Figure \ref{fig:boat2}h.
The lack of bending means no aggregation is observed {(Figure \ref{fig:boat3}a and Supplementary Video \cref{vid:S5})}.

Aggregation kinetics vary at different times.
At higher temperatures, aggregation time scale was not apparently different between different temperatures (or coil shape). 
The \# of clusters decreases to below 25\% of the initial number of ribbons within the first 20 s (early times) for all systems. 
Following the steep decline, aggregation occurs more slowly, eventually tending to an apparent steady state within 100 s (late times). 
During this time, collisions continue to happen between the formed clusters leading to the emergence of two main behaviors.
The first behavior includes systems that did not achieve a steady state \# of clusters smaller than $10$\% of initial units.
Such systems, e.g., $\theta=10^\circ,$  $T=120^\circ$C, exhibited repeated attachment and detachment events tending to a fluctuation around the late-time value, Figure \ref{fig:boat3}b.
The second behavior includes systems that achieved a steady state \# of clusters smaller than $10$\% of initial units, favoring entanglement into large clusters.
In these systems, the collisions resulted in the coalescence of the clusters through interactions between the dangling edges of different clusters, or the annexation of individual ribbons toward subsequent cluster growth, $\theta=10^\circ,\,T=65^\circ$C in Figure \ref{fig:boat3}a.

A low number of resulting clusters provides an indication of `successful' entanglement.
However, to determine which of the above scenarios occurred, it is important to account for the distribution of cluster sizes.
To account for this measure, we can consider the average number of ribbons in each cluster.
The weighted average cluster size in the experiments presented in Figure \ref{fig:boat3}b ($\theta=10^\circ,\,T=[65,90,120]^\circ$C) is presented in Figure \ref{fig:boat3}c. 
The $\theta=10^\circ,\,T=90^\circ$C ribbon system exhibited a late-time value of $4 \pm 2$ clusters which had an average cluster size of $18 \pm 5$ ribbons/cluster, this constitutes a yield of merely $45\%$ of the initial number of ribbons.
The $\theta=10^\circ,\,T=65^\circ$C exhibited a late-time value of $3 \pm 1$ clusters which had an average cluster size of $34 \pm 4$ ribbons/cluster, constituting a $85\%$ yield of the initial number of ribbons.
This is also evident from time lapse images presented at the rightmost column ($t=120$ s) of Figure \ref{fig:boat3}a.
 {Close-up images of $\theta=10^\circ$ aggregates entangled at different temperatures, $T=[65,90,120]^\circ$C, are presented in Supplementary Figure S6a.}
Thus, a cluster size approaching the initial number of units will indicate a successful entanglement that has resulted in the formation of a single mass.
 {The field rotation speed was shown to affect aggregation quality with rotation speeds below 1.2 Hz resulting in less tightly packed aggregates with reduced mechanical properties, probably due to the lack to kinetic energy required to overcome the initial friction between ribbons to rearrange after initial contact, Supplementary Figure S5.} 


Ribbons with moderate curvature form larger aggregates.
At $65^\circ$C, the average number of ribbons per cluster is greatest with an average yield of approximately 85\%.
This indicates that the moderate amount of curvature exhibited by the ribbons at moderate temperatures  is more favorable for promoting entanglement as compared to the high curvature, tightly coiled shapes found at higher temperatures near $T_{NI}$.
We expect that the relatively low curvature enables interaction that results in the interlocking of the ribbons.
Additionally, the tightly coiled high temperatures conformations at $90^\circ$C and  $120^\circ$C do not supply the formed aggregates with `loose' edges that facilitate growth through coalescence with other clusters or ribbon annexation.



The shape of ribbons at a common temperature resulting from varying the offset angle is also expected to affect aggregate formation.
 {We consider three different actuated shapes based on varying the offset angle, $\theta = [0^\circ,10^\circ,20^\circ]$, exhibiting different amounts of curvature and twist (Supplementary Figure S2).}
As a control, we consider a polydomain ribbon, which does not exhibit any actuation (i.e., remains flat) at elevated temperature.
Time lapse images of dispersions of ribbons with different shapes at 65$^\circ$C are shown in Figure \ref{fig:boat3}d. 
 {Close-up images of aggregates resulting from ribbons with different offset angles that were entangled at 65$^\circ$C, are presented in Supplementary Figure S6b.}
The time evolution of the \# of clusters for different offset angles at 65$^\circ$C is presented in Figure \ref{fig:boat3}e, and the average number of ribbons per cluster is presented in Figure \ref{fig:boat3}f.
 Again, the flapping motion performed by the polydomain ribbons (red line) did not result in any bending from the ribbons  {(Supplementary Video \cref{vid:S5})} which did not result in entanglement.

 The kinetics of aggregation of different ribbon systems that aggregate did not exhibit any discernible differences. 
 Although some variation is apparent, no substantial difference in kinetics or average cluster size was observed between the different angles at 65$^\circ$C.
 In all systems, two aggregation rates were observed, a fast drop in \# of clusters lasting approximately 20 s followed by a slower decrease lasting approximately 80 s.
 In all systems, the formed aggregates occupied a volume approximately 80-90\% smaller than in their dispersed state, Figure S7.
{This indicates that motion in circular patterns does result in successful entanglement.
Similar observations were further reported by Patil et al. \cite{Patil2023} in a computational study of active filaments representing dynamic reversible entanglement in the California Blackworms.
We note that the worm bodies in the prior study frequently switch direction of rotation and that their bodies are highly flexible. By comparison, the ribbons follow a magnetic field rotating in only one direction and are somewhat rigid under the imposed forces.   }
 The evident considerable entanglement achieved in all systems in which the two of our preliminary conditions are fulfilled, i.e., continuous synchronicity with the field rotation and moderate curvature, can be considered to support our preliminary hypothesis.

\subsection*{Mathematical model for dynamic entanglement\label{sec_topological_model}}

 {We use a mathematical model for the ribbon dynamics to identify the governing mechanisms of the aggregation process.}
 {In particular, this model allows us to understand the role of topological and non-topological mechanisms in aggregation. 
Our model treats the ribbons as rigid filaments (based on the observed change in curvature smaller than 10\%) moving in a cylindrical domain with diameter $3.8L$ and height $L$, where $L$ is the length of a ribbon. }
To promote entanglement, the process is started from a random initial condition in which the center of mass of each ribbon is in the same plane. Each ribbon then experiences constant driving torques, analogous to the forces on the magnetic domain in the LCE ribbons, along with an external stochastic drift force.
The ribbons also feel contact forces and torques arising from their interactions \cite{bergou2008discrete, tong2023fully}  (Supplementary Information, Supplementary Video \cref{vid:S6}). {These contact forces typically include a cohesion term which prevents touching ribbons from separating unless their separation force exceeds a critical value.} To understand how topological and non-topological parameters govern the aggregation process, we vary filament shape (Figure \ref{fig:boat6}a; top row and bottom row) and cohesion strength (Figure \ref{fig:boat6}a-c and Figure \ref{fig:boat6}d-f).

Filaments of different shapes exhibit different propensities to entangle with others, and thus help illustrate how topology affects aggregation. Here, two filament shapes are considered: a low curvature shape, corresponding to $\theta=10^\circ$ ribbons at $65^\circ$C (Figure \ref{fig:boat6}a, top row), and a high curvature shape, corresponding to $\theta=10^\circ$ ribbons at $90^\circ$C (Figure \ref{fig:boat6}a, bottom row). When cohesive forces are switched on, and the filaments are `sticky', these different shapes give rise to strikingly different aggregation dynamics (Figure \ref{fig:boat6}b) which resemble the experimental results in Figure \ref{fig:boat3}b.

 {The number of links in each filament provides a topological measure for successful entanglement.
We introduce the excess linking per filament (Figure \ref{fig:boat6}c), which quantifies the entanglements in a multi-filament system.} This quantity is built out of the open linking number, $Lk^O_{ij}$, which captures pairwise entanglement between the $i$'th and $j$'th filament \cite{Patil2023, panagiotou2010linking}. For $N$ filaments, the excess linking per filament is the average amount of linking with a single filament, relative to the average amount of linking found in a random configuration of filaments (Supplementary Information). At $200T_0$, the low curvature filaments exhibit a much more robust amount of entanglement than the high curvature filaments, as expected from their larger cluster sizes (Figure \ref{fig:boat6}b). Previous work has indicated that an open linking number with magnitude $1/2$ approximately matches an intuitive notion of entanglement between two filaments \cite{Patil2023}. Since the high curvature filaments achieve an excess linking per filament of approximately $1/2$ (Figure \ref{fig:boat6}c), this suggests that these filaments effectively pair off into clusters of size 2. This is consistent with the cluster number results (Figure \ref{fig:boat6}b), which show the high curvature filaments forming $\approx N/2$ clusters from $N$ initial filaments.

The time evolution of the excess linking per filament additionally provides information on the end-state of the aggregation process. In particular, this order parameters appears to have a positive late-time slope for all filaments in Figure \ref{fig:boat6}c. This indicates that although the number of clusters is approximately constant beyond $100T_0$, there is a continued, albeit slower, growth in cluster size during this time. Additionally, this rate of growth is much faster for lower curvature filaments, indicating a higher propensity for entanglement, in agreement with the experimental data in Figure \ref{fig:boat3}b and  Figure \ref{fig:boat3}c.

 {Cohesion is required to facilitate aggregate growth through topological interaction.}
In addition to topological entanglements, the non-topological cohesive forces themselves play a pivotal role in the assembly of the aggregate (Figure \ref{fig:boat6}d). A representative time-lapse of simulations for `non-sticky' (Supplementary Information) low and high curvature filaments are shown in Figure \ref{fig:boat6}d (see also Suplementary Video \cref{vid:S6}).
In this scenario, the filaments behave as smooth objects that are devoid of any cohesive interactions.
As a result, their cumulative entanglement is solely due to topological interactions.
Importantly, the time evolution of the \# of clusters in the `non-sticky' case (Figure \ref{fig:boat6}e) is drastically different than that of the `sticky' case (Figure \ref{fig:boat6}b).
At early times ($t<25T_0$),  a very small decrease in the \# of clusters is apparent, however both filament types have cluster numbers above 80\% of their initial number. The system then maintains an approximately constant rate of neutral collisions which do not lead to entanglement-induced aggregation. This picture is further supported by the excess linking per filament (Figure \ref{fig:boat6}f), which approaches zero at late times, indicating the absence of meaningful aggregation. The fact that the excess linking does not reach zero is due to the choice of initial state (Supplementary Information) and indicates that the process has not fully equilibrated. Notably, the large initial spike in the excess linking per filament for low curvature filaments further demonstrates their higher innate propensity for topological entanglement.

 {Filament curvature controls the resulting aggregate size and shape.}
Following an extended simulation time, low curvature filaments form larger aggregates (Figure \ref{fig:boat6}g) than those formed using high curvature filaments (Figure \ref{fig:boat6}h).
 {This is further evident from experimental results presented in Figure \ref{fig:boat3}a, and Supplementary Figure S6a.}
The distribution of the number of contacts for low curvature filaments in an aggregate takes larger values than the contact distribution for high curvature filaments. This again illustrates that in large aggregates, low curvature ribbons can form a larger amount of connections, providing a mechanism for aggregate growth. 
 {Qualitative comparison of the resulting simulated aggregates in Figures  \ref{fig:boat6}g and Figure \ref{fig:boat6}h and close-up experimental images in Supplementary Figure S6a, shows that the increased curvature of the ribbons results in a tighter entangled core, and less loose or branched edges that limit constructive interactions with new ribbons.}
Since the capacity for high curvature filaments to connect saturates faster, continued collisions may result in non-constructive interactions that explain the lower slope in Figure \ref{fig:boat6}c. The above results could be used to further control the self-assembly of clusters with designer geometric and topological properties.

\subsection*{Mechanical properties of the aggregate\label{sec_mechanical_properties}}

The fluid-like dispersions of ribbons aggregate into solid collectives.
The combination of the high aspect ratio of the ribbon geometry and the induced curvature enable the ribbons to entangle and interlock to form physical bonds.
The formation of such bonds drives a transition from a liquid dispersion into a solid mass.
To confirm the successful formation of a solid phase, oscillatory strain sweep tests of ribbons in silicone oil were carried out using shear rheometry (Figure \ref{fig:boat4}a and Figure \ref{fig:boat4}b). 
All testing on aggregates was performed following the application of a magnetic field at 1.2 Hz (72 RPM) for 4 minutes from an initially dispersed state ($t=0$ in Figure \ref{fig:boat3}a).
 Both storage and loss moduli increase with increased deformation frequency ($\theta=10^\circ$ aggregates formed at $65^\circ$C, Figure \ref{fig:boat4}c). 
 Such a rate dependent response of the material is indicative of viscoelastic response.
Control measurements were taken in which no ribbons were present in the oil (black lines) and with the ribbons placed in the original dispersed state without the prior activation of the magnetic field (gray lines), Figure \ref{fig:boat4}d.
The oil and the dispersion of unentangled ribbons each behaved as liquids with moduli dramatically lower than the moduli of the entangled aggregates, Figure \ref{fig:boat4}d and Figure \ref{fig:boat1}d. 
We report the moduli of the dispersion normalized to the entire plate geometry, despite the fact that if an aggregate forms, the region that contains ribbons is smaller than the plates.


The aggregates have the properties of viscoelastic solids.
At $65^\circ$C (Figure \ref{fig:boat4}d and Figure \ref{fig:boat4}e) all entangled samples exhibited a viscoelastic behavior that included a linear viscoelastic region at small shear strain values ($\lesssim 0.1\%$) with both storage and loss moduli at least one order of magnitude larger than those measured for the controls.
At higher strain values, a transition to fluid-like behavior is observed  by the increase of the loss modulus over the storage modulus.
This was followed by a sharp decrease of the two moduli tending to the behavior of the pure viscous oil solution, indicating the dissociation of the aggregate.

The storage modulus of the aggregate depends on the actuated ribbon shape.
Namely, ribbons that  bend ($\theta =0^\circ$, blue lines) and ribbons that bend and mildly twist ($\theta =10^\circ$, green lines) resulted in aggregates with larger stiffness relative to aggregates formed from ribbons with a higher twist per length, $\theta =20^\circ$ in red (cf. Figure \ref{fig:boat4}e and Figure \ref{fig:boat2}).
This finding is consistent with measurements taken for LCE ribbon systems formed through static aggregation (i.e., only thermal actuation from an initially overlayed state) performed by Abdelrahman et al. \cite{Abdelrahman2024}.
However, Abdelrahman et al. \cite{Abdelrahman2024} found purely bending ribbons to form slightly stronger aggregates than those that coil, when formed statically.
Our data provide some evidence to support the postulate that coils with a mild twist facilitate better dynamic entanglement by enabling tight assemblies of interlocked ribbons through interweaving rotational motion.
A similar mechanism was also identified by Patil et al. \cite{Patil2023} for the dynamic entanglement of the California Blackworm.

Entanglement at increased temperatures (or tighter coil conformations) reduces the yield stress of the formed aggregates, cf. Figure \ref{fig:boat3}b,c.  
The storage and loss modulus of mildly coiled, $\theta=10^\circ$, ribbons are presented in Figure \ref{fig:boat4}f ($n=3$). 
There is a sharp decrease in both the storage modulus and the yield stress for aggregates formed at for $90^\circ$C (Figure \ref{fig:boat4}g). 
Only weak aggregation is observed for aggregates formed at $120^\circ$C (approximately tending to the behavior of pure oil, Figure \ref{fig:boat4}e). 
This weak aggregation is in agreement with the reduction in aggregate size in Figure \ref{fig:boat3}c.
We expect that tightly coiled individual ribbons do not entangle efficiently.
Thus, collision events result in destructive/neutral effects rather than aggregation.

Heating aggregated samples can increase bond strength within the aggregate.
Aggregates  of ribbons are formed at $65^\circ$C, in the same way as described above. 
The  testing of the aggregate occurs at different temperatures ($\theta=10^\circ$, Figure \ref{fig:boat4}h, $n=3$).
 {The increase in curvature induced by the thermal actuation above the aggregation temperature increases the bond strength and possibly also the number of bonds between ribbons (Supplementary Figure S6).}
The red line indicates samples that were entangled at $65^\circ$C and subsequently cooled down to the crosslinking temperature, $40^\circ$C, in which isolated ribbons return to their flat conformation.
In large volumes of oil, the aggregate flows and partially redisperses. 
When the aggregate is placed in the rheometer and allowed to cool, macroscopic flow was not obvious (Supplementary Video \cref{vid:S7}).
The cooled samples exhibit a similar storage modulus to that obtained at $65^\circ$C with an increased loss modulus throughout the linear viscoelastic region, which implies higher energy dissipation through viscous flow.
Samples tested at $40^\circ$C have a decreased yield strain as compared to samples tested at $65^\circ$C.
Increasing temperature after aggregation resulted in changes to aggregate dimensions ({Supplementary Figure S8) for which gap heights was appropriately adjusted {(section \ref{sec_experimental} for more details)}.
At $90^\circ$C and $120^\circ$C, the aggregates possess higher storage moduli and yield stresses. The increases in moduli and yield stress (Figure \ref{fig:boat4}i) signify increases in both aggregate rigidity and bond strength.
The data from the $120^\circ$C (black line) experiment is only plotted up to the approximate end of the linear viscoelastic region as dissociation was not observed and slipping occurred.

\subsection*{Disassembly of aggregates\label{sec_dissasembly}}

We hypothesized that aggregate disassembly can be induced through magnetic field rotation at high speeds.
However, in moderately high viscosity oil ($\eta = 49\times 10^{-3}$ Pa$\cdot$s), an entangled aggregate that was subjected to increasingly high rotation speeds of the field did not result in the dissociation of the aggregate, top panel in Figure \ref{fig:boat5}a.
Above a critical frequency, magnetic field rotation resulted in vibration of the magnetic domain, indicating that the rotational kinetic energy did not transfer to the ribbon, Supplementary Video \cref{vid:S8}.
 {This loss of synchronicity is due to the critical `step-out' frequency, representing the threshold above which the applied magnetic torque is not strong enough to keep the ribbon synchronized with the field \cite{Mahoney2014,Won2022agile}. 
The step-out frequency can be controlled through changing the magnetic moment applied on the ribbon, e.g., by changing the ribbon magnetization or the magnitude of the external field strength\cite{Mahoney2014}, or by changing the viscosity of the environment \cite{Won2022agile}.}


The viscosity of the oil surrounding the ribbons is critical to isothermal aggregate dissociation. 
The average step-out frequency of a single ribbon ($\theta =10^\circ$) at $65^\circ$C in a room temperature viscosity of $\eta = 49\times 10^{-3}$ Pa$\cdot$s oil was found to be approximately 3.33 Hz (200 RPM), Figure \ref{fig:boat5}b ($n=3$).
However, a 5.6-fold decrease in a room temperature viscosity ($\eta = 8.8\times 10^{-3}$ Pa$\cdot$s) yielded a two-fold increase in the step-out frequency to 6.67 Hz (400 RPM), Figure \ref{fig:boat5}b ($n=3$).
The step-out frequency was taken as the point in which the response curve departs from the line representing where response frequency equals field rotation frequency (dashed line).
When applying increasing field rotation speeds on an entangled aggregate in low viscosity oil ($\eta = 8.8\times 10^{-3}$ Pa$\cdot$s), one minute exposures resulted in an increasing degree of dissociation, bottom panel in Figure \ref{fig:boat5}a.
To quantitatively explore the process kinetics, we tracked the evolution of the \# of clusters in time over one minute.
Similarly to association kinetics, the majority of the transient process ended after approximately 20 s ({Supplementary Figure S9), and the \# of clusters was averaged over the final 30 second ($n=5$).

The critical threshold for dissociation depends on the aggregate bond strength.
The \# of clusters at different magnetic rotation speeds for aggregates formed at $65^\circ$C in $\eta = 8.8\times 10^{-3}$ Pa$\cdot$s oil and heated to three different temperatures, $T=[65,90,120]^\circ$C is presented in  Figure \ref{fig:boat5}c ($n=3$).
The curves indicate that dissociation into clusters becomes substantial only above approximately 3.33 Hz (200 RPM) at $65^\circ$C and that the amount of dissociation increases with the increase in supplied rotational energy.
An approximate plateau is reached around 6.67 Hz (400 RPM) (i.e., the step-out frequency obtained for $\eta = 8.8\times 10^{-3}$ Pa$\cdot$s).
 {Additional estimation of the rotational Reynolds in the new range, taking into account the reduced viscosity and the increased rotation speeds, provides that rotational Reynolds values increase up to approximately, $Re_\omega \approx 41.5$, considered to be close to the approximate limit for laminar flow\cite{CHILDS2011177}, i.e., 40-60.}

Aggregates entangled at $65^\circ$C but tested at increased temperatures show a shift in the magnetic field rotation speed required for cluster dissociation.
This likely represents an increase in the energetic barrier for dissociation, as represented by the schematic in Figure \ref{fig:boat5}d.
For example, at $90^\circ$C, the minimal frequency for dissociation to occur was approximately 5.3 Hz (320 RPM), and for $120^\circ$C there was no effective dissociation, indicating that the critical barrier for dissociation is again higher than provided below the given step-out frequency of 6.67 Hz (400 RPM).


\section*{DISCUSSION\label{sec_conclusion}}

We report a method to create macroscopic aggregates capable of autonomous assembly and disassembly.
Liquid crystal elastomer ribbons are programmed to transition reversibly from flat to three dimensional shapes combining bending and twisting.
A magnetic domain on the ribbon induces motion through an area much larger than the total length of the ribbon in response to an external, rotating, and uniform magnetic field.
As the ribbons move, aggregation from dispersed states is observed. 
Repeated interactions create entangled clusters that grow over time.
The aggregates retain their structure when the magnetic field is removed.

The governing mechanisms that control ribbon interactions, are studied using a mathematical model that explores their emergent collective topological dynamics.
 {The model supports experimental evidence that the shape of the filaments affects the final size of the aggregate and demonstrates that cohesive forces play a pivotal role in the topological assembly of the aggregate.
Namely, lower curvature filaments have a higher propensity for topological entanglement and are capable of forming larger amounts of interconnections than the tighter, higher curvature filaments. 
This propensity, however, requires cohesion or friction to avoid the ribbons from slipping past each other as they interact with other ribbons in the suspension.
One interpretation may be that the moderate curvature of the ribbons is not sufficient to induce frustration in the ribbons, at least in the early stages of the aggregate formation or at a low number of contacts.
The cohesion thus allows the ribbons to be locked-in and facilitate growth.}

The resulting aggregates behave as viscoelastic solids.
The formed aggregates are shown to exhibit the properties of viscoelastic solids that are able to store and dissipate energy, and the shape of the individual unit is shown to affect both the aggregation dynamics and the properties of the resulting aggregate.
Additionally, we induce controlled dissociation through imparting kinetic rotational energy to the individual ribbons at high field rotation speeds. 
Ribbon shape and the medium in which dissociation occurs govern disassembly. 
In essence, slower motion of the ribbons enables the constructive entanglement of the ribbons, while higher velocity motion interferes with these processes and results in dissociation.
 {In light of the mechanistic results from the model, the barrier for dissociation can be assumed to be related to the frictional energy between the aggregated ribbons.}

The individual motion apparent in both the constructive (assembly) and deconstructive (disassembly) mechanisms are reminiscent of the dynamic behaviors prevalent in aggregating systems in the animal kingdom, such as worms and fire ants. 
 Imparting such abilities into synthetic systems may enable a wide range of materials to be assembled on demand, enabling a range of potential applications from bio-inspired soft robotics to injectable biomaterials.

\subsection*{Limitations of the study and future outlook}


 {This work presents a method for assembling solid aggregates from dilute suspensions of responsive shape-changing polymer ribbons.
A previous work by Abdelrahman et al. \cite{Abdelrahman2024} demonstrated the assembly of similar polymer ribbons from concentrated suspensions  by changing temperature.
This temperature change induces morphing in the individual ribbons from flat to bent and twisted shapes.
These curvatures form topological interactions that act as physical bonds that provide structural stability to the aggregate, providing it with solid-like properties.
However, since motion arises solely due to the actuation of the ribbon, such interactions require close proximity between the ribbons (smaller than half a ribbon length), i.e., high concentrations.
Herein, we added an additional mechanism for motion, induced by an external, rotating magnetic field.
This additional motion allows the separation of the mechanism that controls the shape of a ribbon from the mechanism that drives the interaction of the individual ribbons. 
These separate mechanisms enable aggregates to form with solid-like properties at constant temperatures and low starting concentrations. 
Fundamentally, this achievement overcomes the limitation on distance between ribbons to lengths much larger than the length of a single unit \cite{Abdelrahman2024}.
However, we note that the current method is limited to the region in which the field is uniform (approximately 60 mm diameter circle, Supplementary Figure S3).
Such limitations may prevail through other external sources of propulsion, whereas internal propelling forces \cite{Sanchez2012, Marcy2004, Tavera2024} may completely eliminate this limitation.
Future studies may seek to study the effects of such alternative mechanisms.

Magnetic-field induced motion opens the way for advanced system manipulation.
Imparting the individual ribbons with magnetic responsiveness allows for variation in ribbon response by modulation of the magnetic field. 
This work has demonstrated that change in the rotation frequency allows for transition between assembly and disassembly dynamics between ribbons. 
Additional progressions of this work may include the individual control of different ribbon populations that have different magnetization values, resulting in different step-out frequencies and different dipole moments or magnetic torques under an applied field.
For example, Nelson et al. \cite{Mahoney2014} demonstrated that systems of rotating magnetic microrobots can be controlled simultaneously when operated above the step-out frequency.
Wie et al. \cite{Won2019} used modulation of the magnetic field that induce centrifugal forces to control micro-robots, an approach that has the potential to contribute additional control to the ribbon dynamics.
Such concepts, along with careful design of the geometry of the magnetically responsive unit and modal manipulations of the applied magnetic field, have been shown to result in variable collective task-execution and self-organization \cite{Xie2019,Won2022agile,Yang2024}.

The assembly of aggregates in a fluid medium requires a better understanding of hydrodynamic effects.
This work has demonstrated the important contribution of the fluid medium on the dynamics of individual ribbon motion and disassembly of the collective.
Expectantly, the induced motion of the individual ribbons and ribbon clusters induces flow.
This was briefly demonstrated in this work using simple flow visualization in different scenarios (Supplementary Video \cref{vid:S9}).
This visualization allows us to eliminate the occurrence of large-scale circulatory flows that act as sinks, resulting in ribbon clustering.
Additionally, the fluid does not move the ribbons once the field is stopped. 
Therefore, we have postulated that under the assembly conditions of this study, hydrodynamics play a lesser role as compared to ribbon form.
There has been significant discussion in the literature on the effect of a compliant filament in different flow fields \cite{becker2001instability,liu2018morphological}.
Additionally, high rotation speeds may result in local turbulence that can affect the dynamics of dissociation.
Future studies may choose to employ flow measurement techniques to elucidate such phenomena.

Linking between ribbons relies on cohesive interactions.
The mathematical model presented in this work demonstrated that the successful formation of topological links between curved ribbons relies on the existence of forces between the surfaces of the ribbons. The contact force law considered here is a model for adhesion, where contacts become pinned until they are separated by a sufficiently large force. However, there are several other examples of contact laws that incorporate other effects, such as sliding contacts, which are not captured here. Understanding how aggregation is affected by these different contact force protocols, such as static friction, represents an important future challenge. 
It may also be inferred that for aggregate dissociation to occur, such cohesive interactions must be overcome.
However, this study did not discern the energy of these  interactions and the effect its modulation would exert on the resulting aggregate shape and size.
{Quantitative information on the nature of interactions within the aggregates requires three-dimensional imaging capabilities such as micro-CT \cite{Abdelrahman2024} or ultrasound \cite{Patil2023}.}
Future work may explore a more in-depth characterization and estimation of the work required to separate interlocked ribbons, for example as a function of variable surface roughness or  contact surface area.

Ribbon compliance and dimensions can significantly affect the behavior of the system.
The use of polymer ribbons for aggregation as presented in this study exists in a vast parameter space.
Two important parameters that have been fixed in this study were the crosslink density (affecting its elastic modulus) and the dimensions of the ribbons.
These parameters act to change the effective ribbon stiffness.
Namely, in this work we studied the ribbons in the rigid limit, where the magnetic field induces mostly rigid body rotations rather the bending moments.
Relaxing this limitation,  with lower modulus or longer/thinner ribbons, is expected to substantially change the mechanisms of interaction, the importance of hydrodynamic effects, and the resulting aggregate properties, and mechanical response.
Future studies realizing such flexible and highly compliant filaments are expected to unlock unique behaviors.}

\newpage

\section*{EXPERIMENTAL PROCEDURES\label{sec_experimental}}

\subsection*{Materials} 
Liquid crystal monomer 1,4-bis-[4-(6-acryloyloxyhexyloxy)benzoyloxy]-2-methylbenzene (RM82) was purchased from Henan Daken Chemical Co.,Ltd and monomer  1,4-bis-[4-(3-\\ acryloyloxypropyloxy)benzoyloxy]-2-methylbenzene (RM257) was purchased from Wilshire inc. KG. Photoinitiators Irgacure I-369 and Irgacure I-651, chain extender n-butylamine, dimethylformamide, toluene,  acetone, isopropanol, silicone oil ($49$ mPa$\cdot$s) were purchased from Fisher Scientific. Low viscosity silicone oil ($8.8$ mPa$\cdot$s) was purchased from Clearco inc. Brilliant yellow dye was purchased from Chem-Impex inc. Neodymium-iron-boron (NdFeB) magnetic microparticles (average diameter = 25 µm) were purchased from Magnequench. A two-part polydimethylsiloxane (PDMS) elastomer was purchased from Ellsworth Adhesives. High temperature block magnets (BX088SH) were purchased from k\&J Magnetics. Glass microspheres (\#22) were purchased from Fibre Glast.

\noindent\subsection*{Synthesis and Preparation\label{sec_fab}} \vspace{-6pt}
Clean glass slides (VWR) with dimensions of 38mm$\times$75mm$\times$1mm were subjected to O$_2$ plasma for 5 min. Subsequently, the slides were spin-coated with 1wt\% brilliant yellow solution in dimethylformamide at a spin speed of 1,500 RPM for 30 s. 
The slides were baked for 30 min at 100°C on a hotplate to ensure the evaporation of dimethylformamide.
Each coated glass slide was exposed to linearly polarized broadband visible light from a modified projector at an intensity of 10 mW/cm2 for 7 min to align the brilliant yellow molecules \cite{Boothby2017}. 
For the preparation of polydomain samples, no polarized light exposure was used.
After alignment, the coated glass slides were spin-coated at a speed of 1,000 RPM for 15 s with a 9 wt\% RM257 (reactive mesogens) and 1 wt\% I-651 (photoinitiator) toluene solution (10 wt\% solids). 
The slides were then irradiated with 365 nm ultraviolet (UV) light at an intensity of 2 mW/cm$^{–2}$ for 5 min using a UV oven (UVP Crosslinker CL-3000) for polymerization. 
Glass cells were prepared by adhering a pair of RM257-coated glass slides together using a 50 µm Kapton Polyimide film spacer (McMaster Carr) and superglue (Gorilla). 
A 1.4:1.0 molar ratio of RM82 to n-butylamine was heated and mixed with 1 wt\% of I-369.
The solution was then allowed to partially fill the glass cell through capillary action on a hotplate at 75 °C before placing it in an oven at 75 °C to complete the filling and oligomerization.
After 15 h of oligomerization, the glass cells were allowed to cool to 40 $^\circ$C temperature before exposing the cell to 365 nm UV light (Lumen Dynamics, OmniCure LX400+) at an intensity of 12 mW/cm$^{–2}$ for polymerization for 50 min on each side. 
After polymerization, an additional slide coated with brilliant yellow and polymerized RM257 (as mentioned above) was attached to the cell in the plane of one of the glass slides to create space for the magnetic domain.
The slide was separated  using a 100 µm Kapton Polyimide film spacer (McMaster Carr) and adhered to the cell using superglue.
The formed space was filled with a 1.4:1.0 molar ratio of RM82 to n-butylamine with 20 wt\% (NdFeB) magnetic microparticles and 1 wt\% of I-369.
An additional oligomerization cycle was performed as mentioned above before exposure to 365 nm UV light (Lumen Dynamics, OmniCure LX400+) at 40 $^\circ$C under an intensity of 12 mW/cm$^{–2}$ for polymerization for 15 min on each side. 
After polymerization, the top glass slide was released, and the film was cut using a CO$_2$ laser cutter (Gravograph CO$_2$ laser LS100-40W). 
The cut ribbons were washed in water to remove brilliant yellow residue and the magnetic domains were magnetized under a 2.8 T impulse magnetic field (ACS Scientific IM-10-30).
The twisted nematic ribbons were magnetized at an angle corresponding to the offset angle with respect to the long axis of the ribbon.
Polydomain ribbons were magnetized in the direction parallel to the long axis of the ribbon.
\noindent\subsection*{Temperature controlled rotating magnetic field apparatus} \vspace{-6pt}
The apparatus frame was made from aluminium and constructed using a custom design.
A uniform magnetic field was achieved by forming a circular Halbach array (Halbach cylinder, $k=2$) made from an aluminium frame housing 16 Nickel-coated high temperature block magnets (N42SH) with dimensions 1"$\times$1/2"$\times$1/2" arranged in a circle with a 115 mm diameter, Figure S3. 
The array was mounted on a 24V brushed DC Electric Motor.
The bottom side of a thin aluminum plate with dimensions 150 mm$\times$150 mm$\times$2 mm was mounted with a 6" adhesive silicone rubber heater 120 VAC, 10 W/in$^{2}$ (35765K423, McMaster-Carr).
 {The plate was mounted 20 mm above the Halbach array, resulting in an average field strength of approximately 26 mT in the plane of the plate, with good uniformity inside the area of a 60 mm ring (Figure S3)}.
The heater was connected to a Process PID controller (48VFL, EXTECH) powered by a variable AC voltage Transformer (TDGC-0.5KM, VEVOR) set to 94 V. 
The oil temperature was checked using a thermocouple (20250-91, Digi-Sense). 

\noindent\subsection*{Magnetic Characterization} \vspace{-6pt}
 { The hysteresis loop of the magnetized LCE samples was measured with a vibrating sample magnetometer (VSM) (Quantum Design MPMS3) ranging from $-3$ T to $3$ T. 
 Three batches of composite LCE samples with 20wt\% magnetic  particles were synthesized and molded into flat films (using the method described above).
 The samples were folded to fit into the VSM sample holder. The magnetic moment of each sample was measured when the applied magnetic field was zero. The remanent magnetization $M$ of the composite was calculated by normalizing the measurement of magnetic moment to the sample weight ($n = 3$).}

\noindent\subsection*{Curvature and twist measurements} \vspace{-6pt}
 Curvature and degree of twist per unit length were measure by fixing a small (5 mm) piece of ribbon in place using a clip submerged in a temperature controlled oil bath.
 The induced curvature at different temperatures was evaluated by fitting a circle to the curved surface of the actuated ribbon.
 The degree of twist was evaluated by tracking the angle between the free end and the fixed end of the ribbon.
 Three separate batches of ribbons were tested for each condition.
\noindent\subsection*{Video tracking and analysis} \vspace{-6pt}
Videos of experiments were captured using a Canon Rebel T5i camera from top view.
Tracking of the magnetic domain was conducted on 20 s videos taken at 60 FPS using a custom Matlab code (Supplementary Information).
The code captured the $x(t)$ and $y(t)$ coordinates of the average center of the magnetic domain in time.
To simplify the result and elucidate the features directly related to the magnetic field, the obtained time series were filtered using Matlab's `Bandpass' subroutine in the range [0.5,1.4] Hz to attenuate lateral motions that occur on different time scales than that of the field rotation, set to 1.2 Hz (72 RPM).
Aggregation experiments were performed by capturing 2 min videos taken at 60 FPS. Analysis was applied on 40 frames taken at equal 3 s intervals by manual sampling of individual clusters and ribbons.
Dissociation experiments were performed by capturing 1 min videos taken at 60 FPS. Analysis was applied on 10 frames taken at equal 6 s intervals by manual sampling of individual clusters and ribbons.
To account for the effect of the viscosity, i.e., 'step-out' frequency, the motion of a single ribbon in the field was tracked in time at different field rotation speeds.
Then, the power spectrum of the signal was obtained using the Matlab `pspectrum' subroutine, and the largest non-zero frequency peak value was considered for the rotational frequency response of the ribbon.  
Three separate batches of ribbons were tested for each condition.

\noindent\subsection*{Rheological characterization} \vspace{-6pt}
To measure the rheological properties of aggregates or dispersions, previously reported experiments were followed with modification \cite{Abdelrahman2024}. 
Briefly, parallel-plate rheological scans were conducted utilizing a rheometer (Anton Paar Physica MCR 301). 
An aluminium well (50 mm diameter) was placed on the base to contain the ribbons and silicone oil in the rheometer (Figure \ref{fig:boat4}). 
The oil temperature was checked using a thermocouple (20250-91, Digi-Sense). 
A consistent gap height of 3.5 mm was validated using the controlled rheometer displacement.
Control experiments were performed in which no ribbons were placed in the oil (labeled `Only oil') and in which the ribbons were dispersed in the way similar to the initial condition of the aggregation expeirments (labeled `No field'), Figure \ref{fig:boat4}b.
Aggregates were formed by dispersing the ribbons in a dish and applying the field rotation at 1.2 Hz (72 RPM) for 4 minutes.
Then, the formed aggregate was manually transferred using tweezers to the preheated aluminum well mounted on the rheometer.

Oscillatory strain and frequency sweeps with a gap of 3.5 mm were performed for measuring the storage and loss moduli as a function of shear strain from 0.01\% to 100.00\% (at an angular frequency of 10 rad/s$^{–1}$), and 0.05 rad/s to 100 rad/s (at 0.05\% strain). 
Measurements taken at 90 $^\circ$C were taken using a 3.35 mm gap height, approximately 4.5\% lower than 3.5 mm, based on the average change in aggregate height associated with heating (Supplementary Figure S8).
Measurements taken at 120 $^\circ$C were taken using a 3.18 mm gap height, approximately 9\% lower than 3.5 mm, based on the average change in aggregate height associated with heating (Supplementary Figure S8).
The average changes in aggregate heights were obtained by measuring the dimension change relative to a submerged reference in a controlled oil heat bath.
The yield stress and yield strain values were measured as the shear stress and strain value at the cross-over point between the loss modulus and storage modulus \cite{Perge2014,Dinkgreve2016}. 
Three separate batches of ribbons were tested for each condition.

\noindent\subsection*{Grazing-incidence wide-angle X-ray scattering} \vspace{-6pt}
Grazing-incidence wide-angle X-ray scattering (GIWAXS) was performed at the BioPACIFIC MIP user facilities at UC Santa Barbara. 
GIWAXS was used to evaluate the alignment of  LCE films.
Measurements were taken from homogeneously aligned (monodomain) nematic LCE films produced in the exact same way as described above, with the exception that the nematic director was uniform along the thickness of the film.
Grazing incidence experiments  enable scattering measurements through the surface of a thin films, for which similar transmission geometry experiments would be cumbersome.
As the bottom half of the diffraction cone is absorbed by the substrate, only the upper half of the scattering pattern (Debye-Scherrer ring) is apparent (Supplementary Figure S1)  \cite{Mahmood2020}.
The WAXS instrument was custom-built using a high brightness liquid metal jet X-ray source (Excillum MetalJet D2+ 70 keV) and a 4-megapixel hybrid photon-counting X-ray area detector (Dectris Eiger2 R 4M). 
WAXS data were collected using a 9.24 keV beam in 10-minute intervals at an incidence angle of approximately 1$^\circ$ at room temperature.

\subsection*{Resource availability}


\subsubsection*{Lead contact}


Requests for further information and resources should be directed to and will be fulfilled by the lead contact, Taylor H. Ware (taylor.ware@tamu.edu).

\subsubsection*{Materials availability}

  This study did not generate new materials.

\subsubsection*{Data and code availability}





\begin{itemize}
    \item The data reported in this paper is available from the lead contact upon request.
    \item All original code relating to the mathematical model is publicly available  at  GitHub (\url{github.com/vppatil28/rigid_filaments}) as of the date of  publication.
    All original code related to tracking and analysis of experimental data is attached in the supplementary information of this paper. 
    \item  {The Matlab code used to model the Halbach array \cite{Soltner2010Halbach} is available at\\ \url{https://www.blogs.uni-mainz.de/fb08-physics-halbach-magnets/software-2/}.}
    \item Any additional information required to reanalyze the data reported in this paper is available from the lead contact upon request.
\end{itemize}


\section*{Supplemental information index}




\begin{description}
  \item Videos S1-S9 in mp4 format.
  \item Videos S1-S9 descriptions and their legends in a PDF.
  \item Figures S1-S11 and their legends in a PDF.
  \item Details on mathematical models in a PDF.
  \item Original code used in this work in a PDF.
 
\end{description}

\section*{Acknowledgments}


We thank Justin M. Gonzalez and the Fischer Engineering Design Center team for their assistance with samples preparations.
This work was supported National Science Foundation under Awards No. DMR-2041671 \& CMMI-2427025 and Texas A\&M University. 
This work was partially supported by the BioPACIFIC Materials Innovation Platform of the National Science Foundation under Award No. DMR-1933487.

\section*{Author contributions}


Conceptualization, A.D., T.H.W., and M.K.A.; methodology, A.D., V.P.P., M.S.K., S.T.; investigation, A.D., C.B., K.H., D.C.C., J.A.G., P.A.K., Y.L., and V.P.P.; visualization, A.D., S.M.G., and M.S.K; writing – original draft, A.D. and T.H.W.; writing – review \& editing, A.D., C.B., M.S.K., K.H., S.M.G., D.C.C., J.A.G., S.T., P.A.K., Y.L., M.K.A., V.P.P., and T.H.W.; funding acquisition, T.H.W.; resources, T.H.W, and A.D.; supervision, T.H.W.;

\section*{Declaration of interests}


The authors declare no competing interests.




\bibliography{ref}

\newpage
\section*{MAIN FIGURE TITLES AND LEGENDS}




\begin{figure}[H]
  \includegraphics[width=\linewidth,trim={0 0 0 0},clip]{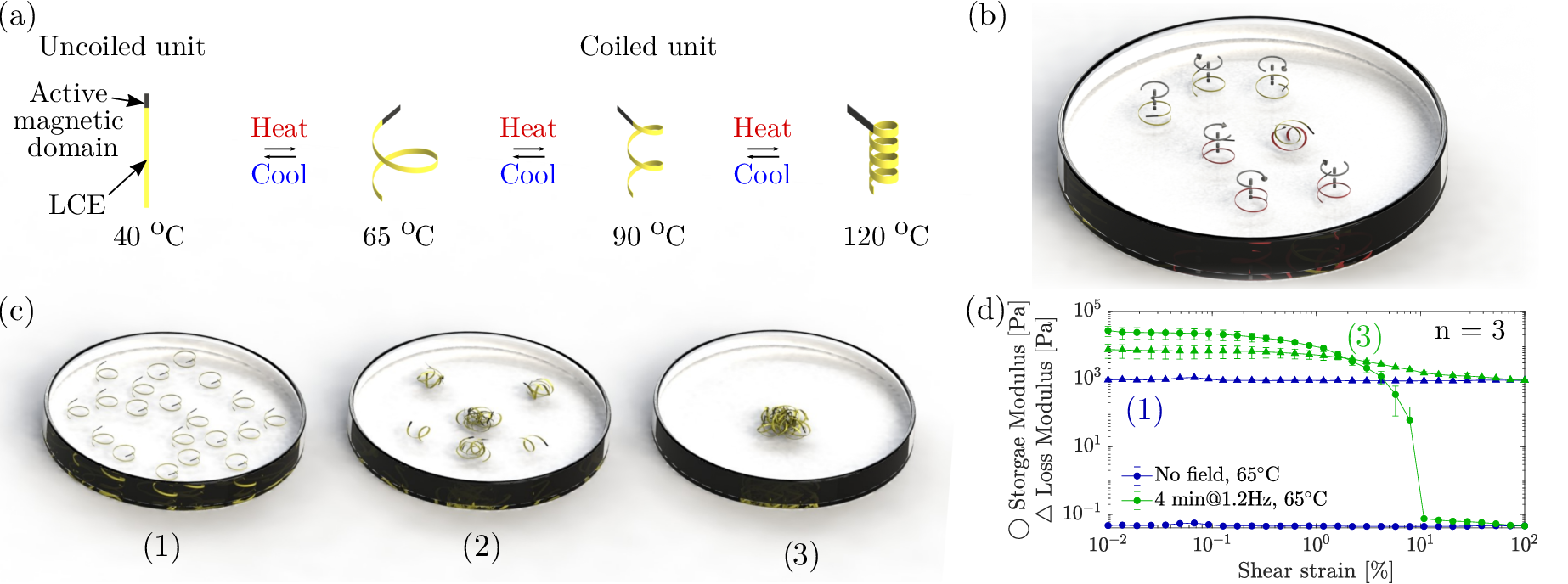}
  \caption{Schematic illustrating the mechanisms that drives shape change and entanglement in magnetically active LCE ribbons.
  (a) An initially straight ribbon undergoes an increasing amount of bending and twisting upon heating as a result of nematic-isotropic transition.
  (b) Schematic of thermally actuated ribbons undergoing axial rotation and lateral motion in a rotating magnetic field that results in entanglement.
  (c) A group of dispersed ribbons (1) undergo entanglement into smaller clusters (2) that eventually connect into a single aggregate (3). 
  (d) Oscillatory rheological strain sweeps plotting the storage (round marker) and loss (triangular marker) modulus as a function of shear strain of ribbons with a length of 16 mm and offset angle $\theta = 10^\circ$ before (1) and after (3) the application of the magnetic field. Error bars represent the standard deviation, $n=3$.}
  \label{fig:boat1}
\end{figure}

\begin{figure}[H]
  \includegraphics[width=\linewidth,trim={0 0cm 0 0},clip]{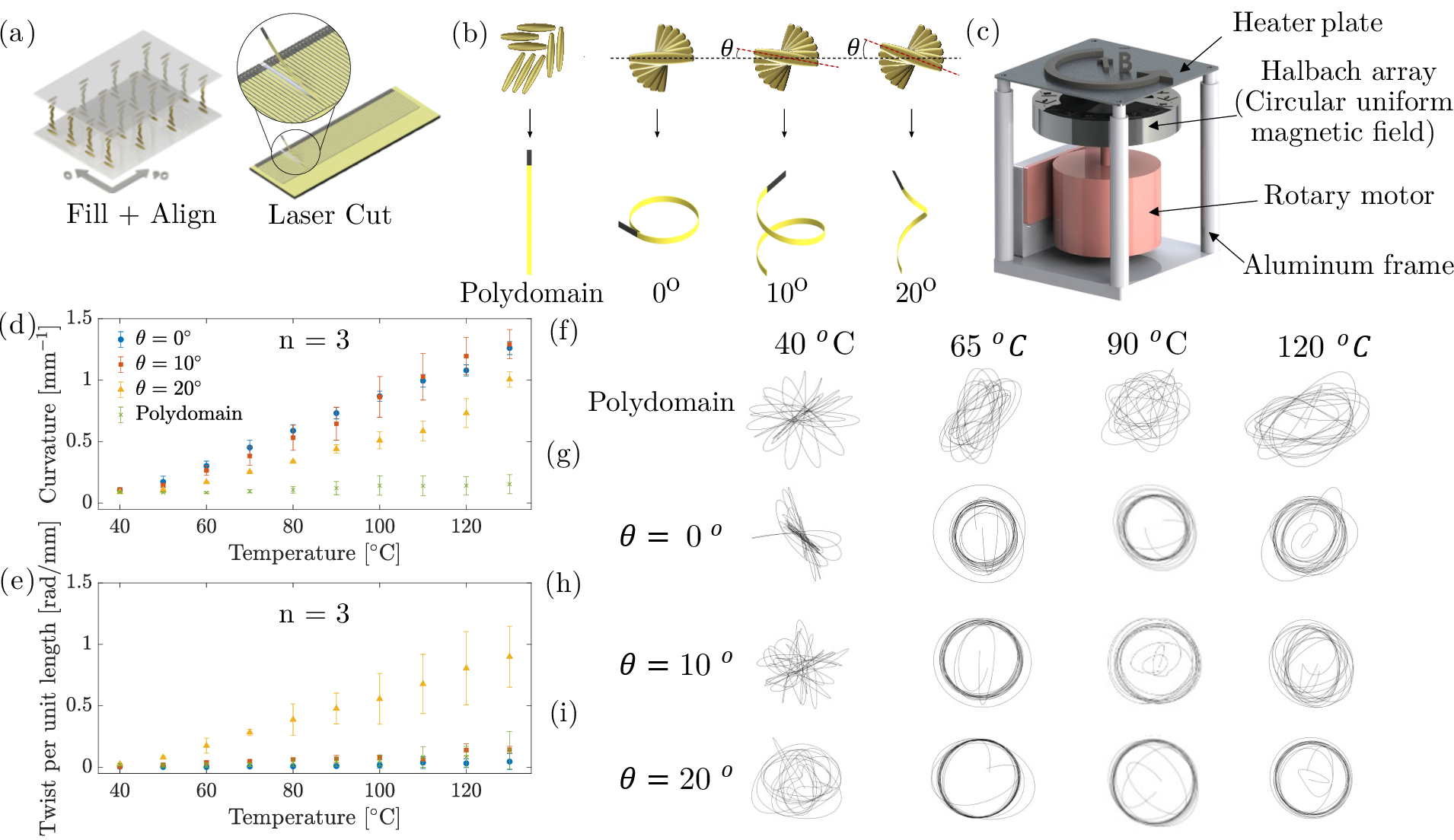}
  \caption{Characterization of thermal and magnetic actuation of single LCE ribbons.
(a) Schematic of the fabrication process of a twisted-nematic LCE ribbon with a magnetic head.
(b) Schematics of actuated LCE ribbons cut at different offset angle, $\theta$, relative to the surface nematic director, resulting in varying amounts of bending and twisting.
(c) Schematic of the temperature controlled rotating uniform magnetic field.
(d) Quantitative characterization of the amount of curvature as a function of temperature for ribbons with different offset angle, $\theta$. 
(e) Quantitative characterization of the amount of twist per unit length as a function of temperature for ribbons with different offset angle, $\theta$. 
(f)-(g) Fourier-filtered motion paths obtained by tracking the trajectory of the active head under a 1.2 Hz (72 RPM) field rotation for polydomain, $\theta = 0^\circ$, $\theta = 10^\circ$, and $\theta = 20^\circ$ ribbons, respectively. In all panels, error bars represent the standard deviation, $n=3$.}
  \label{fig:boat2}
\end{figure}

\begin{figure}[H]
  \includegraphics[width=0.95\linewidth,trim={0cm 0cm 0cm 0cm},clip]{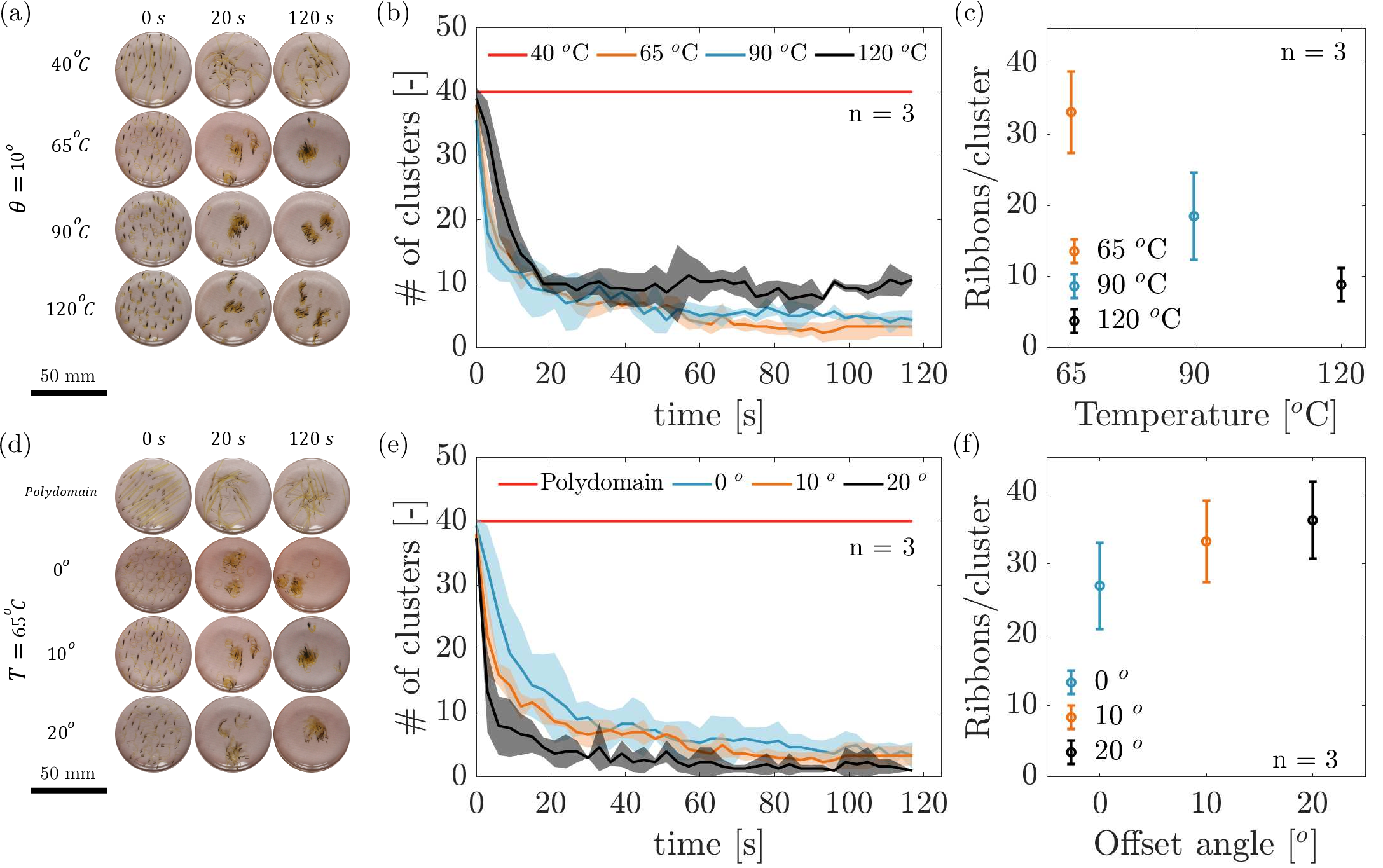}\centering
  \caption{Ribbons entangle based on their shape.
  (a) Time lapse of the entanglement process for different ribbon shapes at different temperatures, taken at $t=[0, 20, 120]$s.  $\theta = 10^\circ$ ribbons at different temperatures. 
  (b) Time evolution of the entanglement process for $\theta = 10^\circ$ ribbons at different temperatures. Shaded region represent the standard deviation, $n=3$.
  (c) Average number of ribbons per cluster after 120 s following the experiment presented in (b). Error bars represent the standard deviation, $n=3$.
  (d) Time lapse of the entanglement process for different ribbon shapes at different temperatures, taken at $t=[0, 20, 120]$s.   Different ribbon shapes at $65^\circ$C based on the fabrication process: polydomain, and offset cutting angle. 
  (e) Time evolution of the entanglement process for ribbons with different offset angle $\theta$ at $T = 65^\circ$C. Shaded region represent the standard deviation, $n=3$.
  (f) Average number of ribbons per cluster after 120 s following the experiment presented in (d). Error bars represent the standard deviation, $n=3$.}
  \label{fig:boat3}
\end{figure}

\begin{figure}[H]
  \includegraphics[width=0.9\textwidth,trim={0cm  0cm 0cm 0cm},clip]{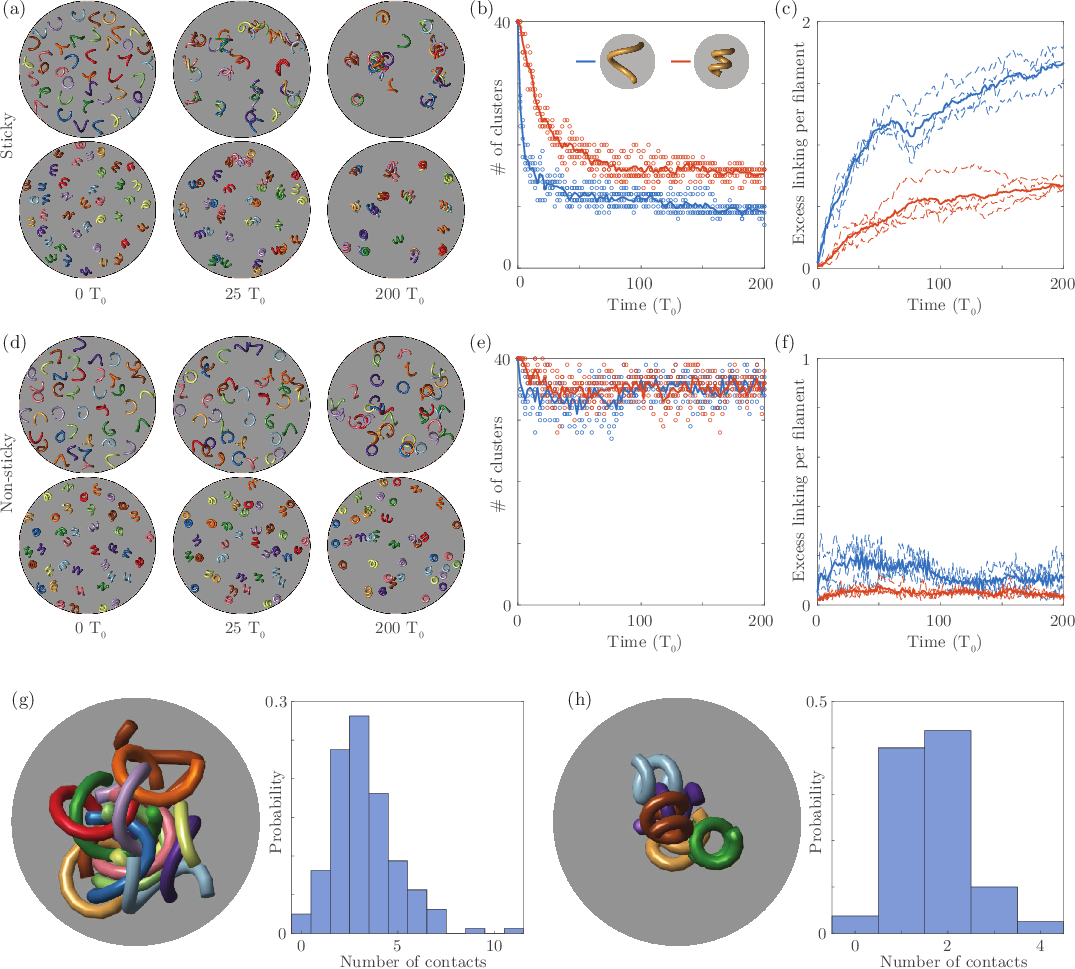}\centering
  \caption{Simulations reveal the topological and non-topological aspects of filament aggregation. 
  (a) Time-lapse images from simulations of 'sticky' filaments  (Supplementary Information) exhibit aggregation, starting from a randomly chosen initial state. Filament shapes correspond to $\theta = 10^\circ$ at 65 $^\circ$C [top row] and 90 $^\circ$C [bottom row].
  (b) Fewer clusters containing larger numbers of filaments are formed by the $65^\circ$ filaments (blue), demonstrating their increased capacity for entanglement relative to the $90^\circ$ filaments (red). Circles show individual simulations, solid curves show mean values ($n=4$).
  (c) Entanglement, as measured by excess linking per filament (Supplementary Information), illustrates the topological distinction between the $65\,^\circ$C and $90\,^\circ$C filaments. Dashed curves show individual simulations, solid curves show mean values ($n=4$).
  (d) Time-lapse images from simulations of `non-sticky' filaments (top row: $65\,^\circ$C; bottom row: $90\,^\circ$C).
  (e) Large clusters do not form for either the $65\,^\circ$C filaments (blue) or the $90\,^\circ$C filaments (red). Circles show individual simulations, solid curves show mean values ($n=4$). 
  (f) Excess linking per filament (Supplementary Information) is close to 0, demonstrating that `non-sticky' filaments suppresses the topological effects of (c). Dashed curves show individual simulations, solid curves show mean values ($n=3$).
  (g) Cluster of  $65\,^\circ$C filaments in the `sticky' case (left). The relatively large observed number of contacts for a given filament at time $200T_0$ in simulation (right) explains the size of the cluster.  
  (h) Cluster of $90\,^\circ$C filaments in the 'sticky' case (left). The small observed number of contacts (right) limits the aggregation process. 
  }
  \label{fig:boat6}
\end{figure}

\begin{figure}[H]
  \includegraphics[width=\linewidth,trim={0 0 0 0},clip]{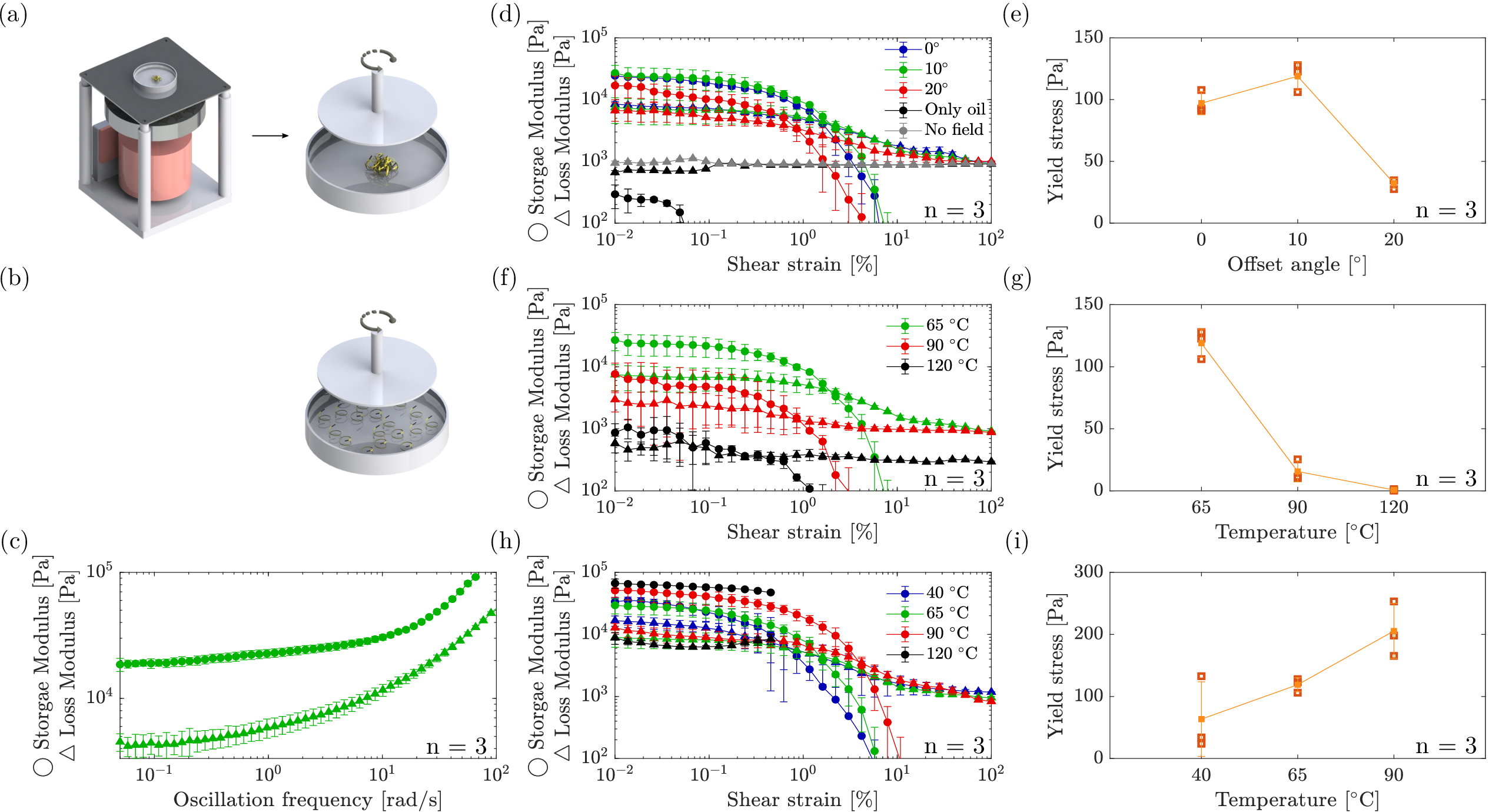}
  \caption{Rheological properties of ribbon aggregates.
  (a) Schematic of experimental procedure. The ribbons are entangled in the magnetic field and are then subjected to a rheological sweep using a parallel plate geometry inside a submerged well of silicone oil.
  (b) Schematic of control procedure. The ribbons are dispersed in a  well of silicone oil without the application of a magnetic field and are then subjected to a rheological sweep using a parallel plate geometry.
  (c) Frequency sweep of entangled aggregates of 40 $\theta = 10^\circ$ ribbons at $T = 65^\circ$C. 
  (d) Strain sweep of entangled aggregates of 40 ribbons with different offset angle $\theta = [0,10,20]^\circ$ at $T = 65^\circ$C.  
  (e) Yield stress values extracted from the respective rheological sweep in (d). 
  (f) Strain sweep of aggregates of 40 $\theta = 10^\circ$ ribbons entangled at different temperatures, $T=[65,90,120]^\circ$C, and tested at the formation temperature. 
  (g) Yield stress values extracted from the respective rheological sweep in (e). 
  (h) Strain sweep of aggregates of 40 $\theta = 10^\circ$ ribbons entangled at $T = 65^\circ$C and tested at different temperatures, $T=[40,65,90,120]^\circ$C. 
  (i) Yield stress values extracted from the respective rheological sweep in (h). In all panels, Error bars represent the standard deviation, $n=3$.
   }
  \label{fig:boat4} 

\end{figure}

\begin{figure}[H]
  \includegraphics[width=0.8\linewidth,,trim={0 0cm 0 0},clip]{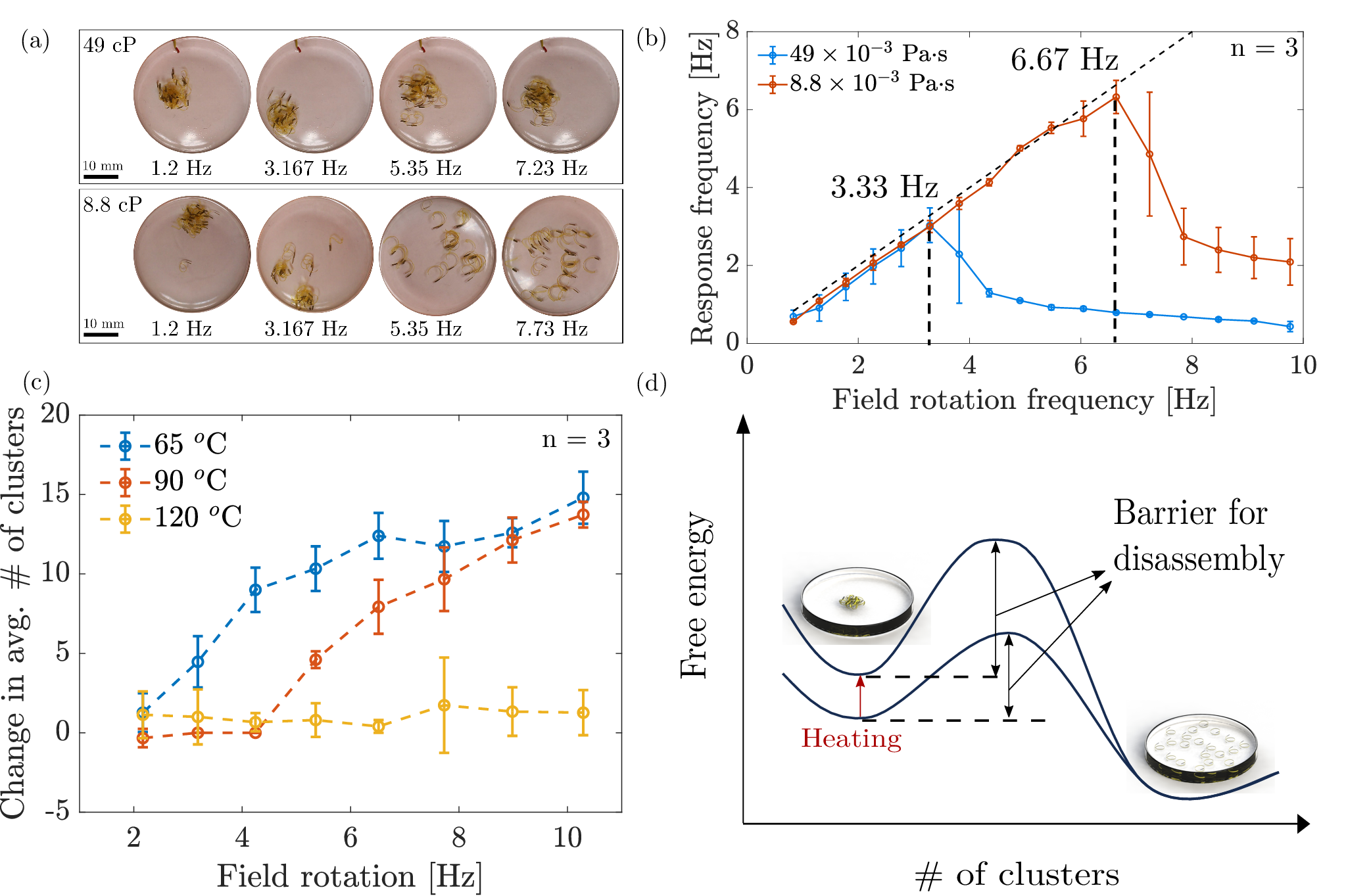}\centering
  \caption{ {Dissociation dynamics of ribbon aggregates.
  (a)  Dissociation of $\theta = 10^\circ$ ribbons at $T = 65^\circ$C in a $49\times 10^{-3}$ Pa$\cdot$s ([top panel]) and $8.8\times 10^{-3}$ Pa$\cdot$s ([bottom panel]) silicone oil subjected to 1 minute magnetic field at different rotation speeds [1.2, 3.167, 5.35, 7.23,    7.73] Hz.
  (b) Rotational response frequency of the magnetic head as a function of different field rotation speeds for silicone oils with different viscosity. (The identified step-out frequency is marked by the dashed lines.) Error bars represent the standard deviation, $n=3$. 
  (c) Change in average number of $\theta = 10^\circ$ ribbons as a function of field rotation at different temperatures, $T=[65,90,120]^\circ$C. Error bars represent the standard deviation, $n=3$. 
  (d) Schematic of the free energy barrier required for dissociation of ribbon aggregates.}
  }
  \label{fig:boat5}
\end{figure}






\bigskip


\newpage

\subsection*{List of supporting videos}

\refstepcounter{vid} \label{vid:S1}
\noindent\textbf{Video  S\thevid: Entanglement and aggregation of ribbons.} \\
 Multi-angle view of aggregation of a ribbon dispersions with $\theta = 10^\circ$ at $65^\circ$C. 

\refstepcounter{vid} \label{vid:S2}
\noindent\textbf{Video S\thevid: Response to the magnetic field.} \\
The trajectories of the magnetic domain of a single $\theta = 10^\circ$ ribbon under different temperatures are tracked in time. The video shows the  the tracked path (left panel), the actual motion of the ribbon (center panel), and the filtered path near 1.2 Hz (right panel) relating to the magnetic rotation.

\refstepcounter{vid} \label{vid:S3}
\noindent\textbf{Video S\thevid: Entanglement and dissociation of four ribbons.} \\
Entanglement of four ribbons ($\theta = 10^\circ$ at $65^\circ$C) in a 30 mm diameter petri dish, followed by cooling to $40^\circ$C and application of the magnetic field.

\refstepcounter{vid} \label{vid:S4}
\noindent\textbf{Video S\thevid: Entanglement and dissociation of forty ribbons.} \\
Entanglement of forty ribbons ($\theta = 10^\circ$ at $65^\circ$C), followed by cooling to $40^\circ$C and application of the magnetic field.

\refstepcounter{vid} \label{vid:S5}
\noindent\textbf{Video S\thevid: Aggregation experiment of different ribbon systems.} \\
Comparison of representative aggregation experiments of: non-actuating ribbons (polydomain); $\theta = [0,10,20]^\circ$ at $65^\circ$C; and $\theta = 10^\circ$ at $T=[40,65.90,120]^\circ$C.

\refstepcounter{vid} \label{vid:S6}
\noindent\textbf{Video S\thevid: Simulations of dynamic entanglement} \\
Represntative simulations of aggregation of `sticky' and `non-sticky' filaments with low ($\theta = 10^\circ$ at $65^\circ$C) and high ($\theta = 10^\circ$ at $90^\circ$C) curvatures.

\refstepcounter{vid} \label{vid:S7}
\noindent\textbf{Video S\thevid: Aggregate response to cooling and heating.} \\
Pre-formed aggregates ($\theta = 10^\circ$ at $65^\circ$C) are cooled to $40^\circ$C and heated to $120^\circ$C.

\refstepcounter{vid} \label{vid:S8}
\noindent\textbf{Video S\thevid: Aggregate dissociation.} \\
Pre-formed aggregates ($\theta = 10^\circ$ at $65^\circ$C) submerged in different viscosity oils are subjected to different field rotation speeds to induce dissociation.

\refstepcounter{vid} \label{vid:S9}
\noindent\textbf{Video S\thevid: Visualization of flow during ribbon motion.} \\
Entanglement of ribbons ($\theta = 10^\circ$ at $65^\circ$C) in an oil solution with 0.75 mg/ml glass microspheres.

\end{document}


\section*{Supplemental information index}




\begin{description}
  \item Videos S1-S9 in mp4 format.
  \item Videos S1-S9 descriptions and their legends in a PDF.
  \item Figures S1-S11 and their legends in a PDF.
  \item Details on mathematical models in a PDF.
  \item Original code used in this work in a PDF.
 
\end{description}

\section*{Supplementary information}
\renewcommand{\figurename}{Figure S\hspace{-2.5pt}}

\setcounter{figure}{0}
\subsection*{Supplementary figures}

\begin{figure}[H]\centering
  \includegraphics[width=0.85\linewidth]{figure_S1.eps}
  \caption{ {Overview of LCE chemistry, and ribbon composition. (a) Drawing of the molecular structure of the monomers (RM82, n-BA), and the photoinitiator (I-369) used to synthesize LCE. (b) Image of a single unactuated ribbon with polarized optical micrographs  between parallel polarizers demonstrating birefringence indicative of a twisted nematic alignment, where  `p' and `a'  denote the polarizer and analyzer orientation, respectively, and 'n' denotes the orientation of the nematic director through the thickness of the film. (c) Magnetic hysteresis curves ($n=3$) recorded for magnetic-LCE composite with 20 wt\% NdFeB magnetic particles. Average remanent magnetization values were extracted from the graph, $M_R = [14.9\pm0.21,-14.5\pm0.56]$ emu/g. (d) 2D grazing-incidence wide-angle X-Ray scattering pattern of a homogenously aligned (monodomain) surface-aligned LCE film. The broad and strong diffraction peak on the meridional (vertical) axis indicates a well-aligned nematic structure. A second set of peaks at lower scattering angles are observed above the equatorial (horizontal) axis. These peaks confirm the presence of smectic ordering. (e) Scattering intensity as a function of azimuthal angle along the respectively colored rings in (d). The vertical trace shows a single nematic peak (blue line), whereas the near equatorial trace shows a strong smectic peak located at $q = 0.18$  \AA$^{-1}$ (where $q$ is the scattering vector, corresponding to an interlayer spacing of 3.49 nm). The fact that the smectic peak is located above the horizontal axis (at approximately 34 degrees) is a strong indication of a tilted smectic phase.    } }
  \label{fig:supp_fab}
\end{figure}

\begin{figure}[H]\centering
  \includegraphics[width=0.90\linewidth]{figure_S2.eps}
  \caption{LCE ribbons with different offset angles, $\theta = [0^\circ,\, 10^\circ,\, 20^\circ]$, at different temperatures, $T = [65,\, 90,\, 120]$ $^\circ$C.}
  \label{fig:supp_ribbon_zoom}
\end{figure}

\begin{figure}[H]\centering
  \includegraphics[width=0.90\linewidth]{figure_S3.eps}
  \caption{ {Design, modeling and validation of uniform magnetic field. (a) Model input of arrangement of magnets along the perimeter ($x-y$ plane) of the Halbach cylinder \cite{Soltner2010Halbach}  ($k=2$) housing 16 block magnets arranged in a circle. (b) Calculated field magnitude the cylinder, $z$, axis. The dotted line represents the height of the the magnets in the array, and the dashed line represents the position of the heater plate relative to the magnetic array. The intersection of the dashed line and the curve indicates the calculated value in the plane of the heater plate. (c) a vector representation of the calculated field direction and magnitude in the center of the magnetic array, i.e., $z=0.0127$ m. (d) A visualization using magnetic particles of the uniform field inside a 60 mm diameter ring located inside the array perimeter on the heater plate. (e) Schematic of the design of rotating circular Halbach array fitted onto the motor.  (f) Spatial measurements of the field in the ring using a Hall sensor.  (g) Tracing of the trajectory of the magnetic domain (of a straight ribbon in water) in time and space for approximately 60 seconds under magnetic field rotation. (h) Analysis of the time series of the position of the domain in time showing an approximately constant velocity profile and the resulting power spectrum showing a single frequency dominates the temporal response of the ribbon.}}
  \label{fig:supp_array}
\end{figure}

\begin{figure}[H]\centering
  \includegraphics[width=0.95\linewidth]{figure_S4.png}
  \caption{Magnetic domain trajectories sampled from different ribbon systems at different temperatures and filtered around 1.2 Hz.}
  \label{fig:supp_paths}
\end{figure}

\begin{figure}[H]\centering
  \includegraphics[width=0.95\linewidth]{figure_S5.eps}
  \caption{ {Characterization of aggregates formed at different rotation speeds. (a) Strain sweep of aggregates of 40 $\theta = 10^\circ$ ribbons entangled at 65$^\circ$C with different magnetic rotation speeds, $\omega = [1.2,0.45,0.25]$ Hz. (b-d) Entangled aggregates after 4 minutes at different magnetic rotation speeds, $\omega = [1.2,0.45,0.25]$ Hz. Aggregate quality below 1.2 Hz is shown to be reduced. } }
  \label{fig:supp_speed}
\end{figure}

\begin{figure}[H]\centering
  \includegraphics[width=0.98\linewidth]{figure_S6.pdf}
  \caption{ {Entangled aggregates. (a) Ribbons with offset angle $\theta = 10^\circ$, formed at different temperatures, $T = [65,\, 90,\, 120]$ $^\circ$C. (b) Ribbons with different offset angles, $\theta = [0^\circ,\, 10^\circ,\, 20^\circ]$, formed at 65$^\circ$C and heated to different temperatures, $T = [65,\, 90,\, 120]$ $^\circ$C.}}
  \label{fig:supp_agg_zoom}
\end{figure}

\begin{figure}[H]\centering
  \includegraphics[width=0.5\linewidth]{figure_S7.eps}
  \caption{Change in absolute occupied volume between dispersed (1) and aggregated (3) states in Figure \ref{fig:boat1}c for three different offset angles, $\theta =[0,10,20]^\circ$. Error bars represent the standard deviation, $n = 3$.}
  \label{fig:supp_vol_change}
\end{figure}

\begin{figure}[H]\centering
  \includegraphics[width=0.4\linewidth]{figure_S8.eps}
  \caption{Average change and standard deviation in aggregate height, relative to the recorded value formed at 65$^\circ$C for $\theta =10^\circ$ ribbons, $n=3$.}
  \label{fig:supp_agg_height}
\end{figure}

\begin{figure}[H]\centering
  \includegraphics[width=0.6\linewidth]{figure_S9.eps}
  \caption{Example of raw data of the time evolution of the \# of clusters in dissociation experiments at different field rotation speeds in the range [130,464] RPM, on a $\theta =10^\circ$ aggregate preformed at 65$^\circ$C and 1.2 Hz.}
  \label{fig:supp_dis_kinetic}
\end{figure}


\newpage
\subsection*{Supplemental Methods}
\subsubsection*{Mathematical model\label{supp_toplogical}}

We formulate a simplified mathematical model for the ribbon aggregation process by treating each ribbon as a rigid body subject to forces and torques. In particular, our model includes the effect of an external driving torque, arising from the magnetic field, a random force accounting for the drift of each ribbon, and contact forces and torques arising from their many-body interactions. The simulations shown in Figure \ref{fig:boat6} are performed by discretizing the ribbons, and then calculating the forces and torques on each ribbon in the discretized framework. Below we describe these forces and torques in more detail.

Treated as a rigid body, the state of a single ribbon can be constructed by rotating a translating a representative ribbon. Specifically, let $\gamma(s)\in \mathbb{R}^3$ for $s\in [0,L]$ be the centerline of the ribbon, where $s$ is an arc length parameter in $L$ is the length of the ribbon. Suppose further that the curve $\gamma$ is centered at the origin, so $\int \gamma \, ds = 0$. The $i$'th ribbon at time $t$, $\gamma_i(s,t)$ can then be described in terms of a $3\times 3$ rotation matrix $Q_i$ and a translation $\bs{R}_i$
\begin{align*}
    \gamma_i(s,t) = Q_i(t) \gamma(s) + \bs{R}_i(t) = \bs{r}_i(t,s) + \bs{R}_i(t)
\end{align*}
Setting $\bs{r}_i(s,t) = Q_i(t)\gamma(s)$, the ribbon velocity decomposes into rotational and translations parts
\begin{align}
    \dot{\gamma}_i = \bs{\omega}_i \times \bs{r}_i + \dot{R}_i
\end{align}
where the angular velocity $\bs{\omega}$ can be written in index notation using the Levi-Civita symbol $\epsilon_{abc}$ as $\epsilon_{abc}\omega_a = \dot{Q}_{bd} Q_{cd}$. The angular velocity therefore satisfies 
\begin{align*}
    \dot{\bs{r}}_i = \dot{Q}_i \gamma = \dot{Q}_i Q_i^{\top} Q_i \gamma = \bs{\omega}_i\times \bs{r}_i
\end{align*}
A state of the $N$-ribbon system can be given in terms of $N$ rotations and translations
\begin{align}
    \{Q_1, \bs{R}_1, Q_2, \bs{R}_2, ... , Q_N, \bs{R}_N\}
\end{align}
An allowed $N$-ribbon configuration must also have the property that no two ribbons intersect each other. To capture contact, we approximate each ribbon as a tube with circular cross-sections of radius $h$ and with centerline $\gamma_i(s,t)$. Using the approximation, an allowed configuration must satisfy the constraint
\begin{align}\label{eq:contact_constraint}
    | \gamma_i(s) - \gamma_j(s') | \geq 2h \qquad \forall \, i,j,s,s'
\end{align}
Additionally, the ribbons are in a domain $\Omega \subset \mathbb{R}^3$. This constraint can be written
\begin{align}
    \gamma_i(s) \in \Omega, \quad d(\gamma_i(s), \partial \Omega) > h \qquad \forall i,s
\end{align}

Each ribbon is subject to an external driving torque $\bs{\tau}^{\text{ext}}$, an external stochastic drift force $\mathbf{F}^{\text{ext}}$, as well as contact forces and torques arising from other ribbons $\mathbf{F}^{\text{con}}, \bs{\tau}^{\text{con}}$ and from the boundary $\mathbf{F}^{\text{bdry}}, \bs{\tau}^{\text{bdry}}$. We treat the dynamics as overdamped and neglect hydrodynamics, obtaining the following simplified equations of motion
\begin{subequations}\label{eq:ribbon_eom}
\begin{align}
    \eta_T \dot{\bs{x}}_i &= \mathbf{F}_i^{\text{ext}} + \mathbf{F}_i^{\text{con}} + \mathbf{F}_i^{\text{bdry}} \\
    \eta_R \bs{\omega}_i &= \bs{\tau}_i^{\text{ext}} + \bs{\tau}_i^{\text{con}} + \bs{\tau}_i^{\text{bdry}}
\end{align}
\end{subequations}
Although this framework corresponds to a simple diagonal resistance matrix, it is sufficient to capture the aggregation dynamics of the ribbons. The external forces and torque do not depend on the state of the ribbon aggregate, and therefore have relatively simple expressions. In particular, we take the external torque to be constant with magnitude $b$
\begin{align}
    \bs{\tau}_i^{\text{ext}} = b \mathbf{e}_z
\end{align}
The external force is stochastic, given by
\begin{subequations}
\begin{align}
    \bs{F}_i^{\text{ext}} &= \cos\theta_i \,\mathbf{e}_x + \sin \theta_i \,\mathbf{e}_y + \sqrt{2D_x}\xi_i^z \,\mathbf{e}_z \\
    \dot{\theta}_i &= \sqrt{2D_\theta} \xi_i^\theta
\end{align}
\end{subequations}
where $\xi_i^z, \xi_i^\theta$ are independent white noise processes satisfying
\begin{align*}
    \langle \xi_i^z(t) \xi_j^z(t')\rangle  = \delta_{ij}\delta(t-t') \qquad \langle \xi_i^\theta(t) \xi_j^\theta(t')\rangle  = \delta_{ij}\delta(t-t')
\end{align*}
The fact that the $z$-direction is treated differently from the $x,y$-directions reflects the fact that the experiments typically take place in domains with horizontal extent much larger than vertical extent.

\begin{figure}[H]\centering
  \includegraphics[width=\linewidth]{figure_S10.pdf}
  \caption{ {Different choices for contact potential (Equations~\ref{eq_heff},\ref{eq_contact_potential}). The contact potential energy $V$ (Equation~\ref{eq_contact_potential}) as a function of $d$ is shown as different values of $h_{\text{eff}}$, along with hard core repulsion.}}
  \label{fig:SI_contact}
\end{figure}

The expressions for contact forces and torques require a more careful accounting of the positions and overlaps of all the ribbons. There are several approaches for formulating these interactions \cite{Patil2023,bergou2008discrete}. Here we construct contact forces by approximating each ribbon as a tube with circular cross-sections of radius $h$ (equation \ref{eq:contact_constraint}). We can quantify the contact between the $i$'th and $j$'th ribbons by setting $\bs{d}_{ij}(s,\sigma) = \gamma_i(s) - \gamma_j(\sigma)$ and $\bs{\hat{d}}_{ij} = \bs{d}_{ij} / | \bs{d}_{ij} |$, and defining $p_{ij}$ as follows
\begin{align}\label{eq_heff}
p_{ij}(s,\sigma) = 1 - \frac{|\bs{d}_{ij}(s,\sigma)|}{2h_{\text{eff}}}
\end{align}
where $h_{\text{eff}} - h$ is small. If $h_{\text{eff}}=h$, then $p_{ij}=0$ when $\gamma_i(s), \gamma_j(\sigma)$ are touching and $p=1$ when $\gamma_i(s), \gamma_j(\sigma)$ overlap completely. Choosing $h_{\text{eff}} \neq h$ has the desirable effect of smoothing out the contact region and the associated contact forces \cite{tong2023fully}. From $p$ we can construct a potential energy $V(p)$ for soft contacts
\begin{align}\label{eq_contact_potential}
V(p) &= \frac{K}{2} p^2 + K p_0 \left( \frac{1}{4} p^4 + \frac{1}{6} p^6 + \frac{1}{8} p^8 \right) \quad \text{for } p>0 \\
V(p) &= 0 \quad \text{for } p<0
\end{align}
where the parameter $K$ sets the magnitude of the repulsive contact forces and $p_0$ is a constant that sets the rate at which the potential stiffens.  {To highlight the effect of choosing different values of $h_{\text{eff}}$, we consider contacts between two chosen points, $\bs{r}_1,\bs{r}_2$ on our set of ribbons. Let $d = |\bs{r}_1 - \bs{r}_2|$ be the distance between these two points. The contact potential energy of these points is given by $V(p)$ where $p = 1 - d/2h_{\text{eff}}$. Since $p$ depends on $d$ and $h_{\text{eff}}$, we can write $V = V(d,h_{\text{eff}})$. Figure ~S\ref{fig:SI_contact} shows plots of $V$ as a function of $d$ for different values of $h_{\text{eff}}$. In each case, contact occurs when $d=2h$. The hard core potential would therefore be infinite for $d<2h$ and zero for $d>2h$ (Figure ~S\ref{fig:SI_contact}, dashed curve). Increasing $h_{\text{eff}}$ has the effect of causing the potential to stiffen more rapidly as $d$ decreases, at the expense of having $V>0$ for $d>2h$ (Figure ~S\ref{fig:SI_contact}, red curve). Here, we choose $h_{\text{eff}}=1.25 h$ to better approximate the hard core potential (Figure ~S\ref{fig:SI_contact}).}

In addition, we assume that ribbons which are in contact must overcome an energy barrier in order to separate. This can be thought of as a `sticky' adhesion force. A ribbon-ribbon contact which is on the point of separating has $p_{ij}(s, \sigma) = 0$ and $\dot{p}_{ij} <0$. The additional adhesion term can therefore be taken to be proportional to $ \delta(p_{ij}) \Theta( - \dot{p}_{ij} ) \, \bs{\hat{d}}_{ij} $ where $\Theta$ is the Heaviside step function. Including adhesion effects, the contact force density on the $i$'th ribbon due the $j$'th ribbon is then
\begin{align*}
\bs{f}_{ij}^{\text{con}}(s) = -\int_0^L d\sigma \, \left(\frac{dV}{dp} \bigg|_{p_{ij}(s,\sigma)} - \alpha \delta(p_{ij}(s,\sigma)) \Theta( - \dot{p}_{ij}(s,\sigma) ) \right) \,  \bs{\hat{d}}_{ij}(s, \sigma)
\end{align*}
where $\alpha$ determines adhesion strength. From this expression we can write down the total contact force and torque on the $i$ ribbon
\begin{subequations}
\begin{align}
    \mathbf{F}_i^{\text{con}} &= \sum_j \int_0^L ds\, \bs{f}_{ij}^{\text{con}}(s) \\
    \bs{\tau}_i^{\text{con}} &= \sum_j \int_0^L ds\, \bs{r}_i(s)\times \bs{f}_{ij}^{\text{con}}(s)
\end{align}
\end{subequations}

Finally boundary forces act to keep each ribbon confined to a domain $\Omega$. The boundary forces act at $\gamma_i(s)$ if $d(\gamma_i(s), \partial \Omega) \leq h$ or equivalent if $\Theta(h -d(\gamma_i(s), \partial \Omega)) =1$. To write down the boundary force, for $\bs{x} \in \Omega$, let $P(\bs{x}) \in \partial \Omega$ be the closest point to $\bs{x}$ on the boundary of $\Omega$. Additionally, for $\bs{y}\in \partial \Omega$, let $\bs{n}(\bs{y})$ be the inward normal to $\partial \Omega$ at $\bs{y}$. Then the boundary force density on the $i$'th ribbon is
\begin{align*}
    \bs{f}_i^{\text{bdry}}(s) = K_b \, \Theta\big(h -d(\gamma_i(s), \partial \Omega)\big) \, \bs{n}\big(P(\gamma_i(s))\big)
\end{align*}
where $K_b$ sets the strength of the boundary force. The boundary forces and torques are then
\begin{subequations}\label{eq:bdry_force}
\begin{align}
    \mathbf{F}_i^{\text{bdry}} &= \int_0^L ds\, \bs{f}_i^{\text{bdry}}(s) \\
    \bs{\tau}_i^{\text{bdry}} &= \int_0^L ds\, \bs{r}_i(s)\times \bs{f}_i^{\text{bdry}}(s) 
\end{align}
\end{subequations}
Our mathematical model for ribbon aggregation is given by equations \eqref{eq:ribbon_eom}-\eqref{eq:bdry_force}.

\subsubsection*{Geometry and topology of ribbon aggregates}

\begin{figure}[H]\centering
  \includegraphics[width=\linewidth]{figure_S11.pdf}
  \caption{ {Aggregation depends on topological and non-topological parameters for modified ribbon shapes.
  Specifically, we consider the ribbon shape with the protrusion of the magnetic domain from the coiled shapes of the ribbons, e.g., Figure S\ref{fig:supp_ribbon_zoom}.
  (a)~Modified ribbon shapes (longer head portions) relative to Figure ~\ref{fig:boat6}, $\theta = 10^\circ$ at 65 $^\circ$C [top panel] and 90 $^\circ$C [bottom panel]. (b-c)~The sticky case (top row) reveals the different aggregation behavior of each filament as a function of curvature. (d-e)~In the absence of adhesion, neither filament forms stable clusters.}}
  \label{fig:SI_aggregation}
\end{figure}

Geometric and topological quantities can be used to understand the structure of the ribbon aggregates formed by the model described above.  {
For example, as a result of the magnetic domain not changing shape, it protrudes from the coiled shapes of the ribbons, e.g., Figure S\ref{fig:supp_ribbon_zoom}. 
The importance of topology in particular can be seen from the fact that modifying the shapes of the filaments (Figure ~S\ref{fig:SI_aggregation}a) relative to the data present in the main text, Figure ~\ref{fig:boat6}, does not affect their topological state (Figure ~S\ref{fig:SI_aggregation}b-e). }
The contact structure of these aggregates follows from constructing the contact graph, where each ribbon is a vertex and two vertices are joined by an edge if the corresponding ribbons are touching. More precisely, let $A_{ij}$ be $1$ if ribbons $i$ and $j$ are touching, and $0$ otherwise. $A_{ij}$ is therefore an adjacency matrix for the ribbon contact graph. The number of connected components of this graph provides a measure of aggregation of the ribbons.

The topological connectivity of ribbon aggregates can be quantified using the open linking number
\begin{align}
    Lk_{ij}^O = \frac{1}{4\pi} \int ds\,d\sigma\, \frac{ (\gamma_i(s) - \gamma_j(\sigma)) \cdot ( \gamma_i'(s) \times \gamma_j'(\sigma) ) }{ |\gamma_i(s) - \gamma_j(\sigma) |^3} 
\end{align}
For closed curves, the above integral yields a topological invariant whose value remains constant as the curves are deformed and moved, provided they do not pass through each other. For open curves, such as the ribbons we study here, this integral varies continuously as the open curves move. However, the open linking number nonetheless provides important information about how intertwined two curves are \cite{Patil2023, panagiotou2010linking}. Furthermore, the open linking numbers also allow us to measure the total amount of linking between the $i$'th ribbon and the rest of the aggregate
\begin{align*}
    T_i^{Lk}(\gamma_1,...\gamma_N) = \sum_j |Lk_{ij}^O|
\end{align*}
The average value of this parameter is the linking per filament, $T^{Lk}$, which gives a measure of the overall total intertwining within an aggregate \cite{Patil2023}
\begin{align*}
    T^{Lk}(\gamma_1,...\gamma_N) = \langle T_i^{Lk} \rangle  = \frac{1}{N} \sum_{i,j} |Lk_{ij}^O|
\end{align*}
A random choice of $N$ ribbons within a domain $\Omega$ will typically have small, nonzero linking per filament $T^{Lk}$. This non-zero `background linking' can be quantified finding the average value of $T^{Lk}$ for random collections of $N$ ribbons. Let $(\tilde{\gamma}_1,...\tilde{\gamma}_N)$ be $N$ ribbons chosen uniformly at random in a domain $\Omega$, subject to the contact and boundary constraints described above (equation \ref{eq:contact_constraint}). The excess linking per filament can then be defined as
\begin{align}
   T^{Lk}_{ex}(\gamma_1,...\gamma_N) =  T^{Lk}(\gamma_1,...\gamma_N) - \langle T^{Lk}(\tilde{\gamma}_1,...\tilde{\gamma}_N)\rangle
\end{align}
To avoid distortions due to the background linking, in Figure \ref{fig:boat6}, we show the excess linking per filament.

\newpage

\bibliography{ref}